\documentclass[%
 aip,
 amsmath,amssymb,
]{revtex4-1}
\usepackage{amsthm}
\usepackage{graphicx}
\usepackage{bm}

\usepackage[utf8]{inputenc}
\usepackage[T1]{fontenc}
\usepackage{mathptmx}
\usepackage{etoolbox}
\usepackage[normalem]{ulem} 

\usepackage[shortlabels]{enumitem}

\def\R{{\Bbb R}}

\usepackage{amsfonts}
\usepackage[utf8]{inputenc}
\usepackage[T1]{fontenc}
\usepackage{mathptmx}
\newtheorem{definition}{Definition}[section]
\newtheorem{thm}{Theorem}[section]

\newtheorem{remark}{Remark}[section]

\newtheorem{lemma}{Lemma}[section]

\usepackage{color}

\usepackage{tikz}
\usetikzlibrary{matrix}

\makeatletter
\def\@email#1#2{%
 \endgroup
 \patchcmd{\titleblock@produce}
  {\frontmatter@RRAPformat}
  {\frontmatter@RRAPformat{\produce@RRAP{*#1\href{mailto:#2}{#2}}}\frontmatter@RRAPformat}
  {}{}
}%
\makeatother
\begin{document}

\preprint{AIP/123-QED}

\title{Diffusion-controlled reaction rate to an active site in a spherical cavity: Extension of Berg's theory} 



\author{Sergey~D.~Traytak} 
\email[Author to whom correspondence should be addressed:]{sergtray@mail.ru}
\affiliation{Semenov Federal Research Center for Chemical Physics, Russian Academy of Sciences, 4 Kosygina St., 119991 Moscow, Russian Federation}
\author{Georgiy~A.~Babushkin} 
\email{georgbabu2912@gmail.com}
\affiliation{Semenov Federal Research Center for Chemical Physics, Russian Academy of Sciences, 4 Kosygina St., 119991 Moscow, Russian Federation}
\affiliation{Moscow Institute of Physics and Technology,  9 Institutskyi Lane, 141701 Dolgoprudnyi, Moscow Region, Russian Federation}



\date{\today}

\begin{abstract}
	This study is due to various applications in physics, chemistry and especially in biology, where both bounded configuration domain and chemical anisotropy could play a great part. In fact we generalize the well-known Berg’s theory, which describes diffusion-controlled reactions occurring within a spherically symmetric absorber-cavity system. The trapping probability and the reaction rate at which a small diffusing particle is captured by an axially symmetric one reactive patch absorber inside a spherical cavity were found semi-analytically and numerically by means of the dual series relations method.
	This approach leads to such incredibly fast convergence, that it may rightly be referred to as exact one. The results obtained can be used to test numerical programmes that describe diffusion-controlled reactions in real physical systems for reactants with arbitrary anisotropic reactivity, which are located inside of various cavities as well as in the unbounded domains. Moreover, we managed to find a close connection between the dual series relations method and the generalized method of separation of variables.
\end{abstract}

\maketitle

\section{Introduction}\label{sec:Intro}

\subsection{Motivation}\label{subsec:Motiv}

One of the main motivation for writing this paper is two-fold: first expand the theoretical range of use of the dual series relations method and secondly outline the  physico-chemical problems, there this theory can be applied. 

The subject of this paper is inspired substantially by two important results obtained previously by H. Berg in Ref.~\onlinecite{Berg93} and by C. Eun in Ref.~\onlinecite{Eun17}.

Howard Berg’s model studied in Chapter 3  "Diffusion to capture" of the classic treatise "Random Walks in Biology",~\cite{Berg93} which served as the handbook for researchers in biology, biochemistry, and biophysics for many years. Note in passing that in our review~\cite{Piazza19} we have already paid tribute to this very chapter.

Berg  started his problem statement 
by saying:"Suppose a particle is released near a spherical adsorber 
of radius $R$ at a point $r = R_0> R$? What is the probability 
that the particle will be adsorbed at $r = R$  rather than 
wander away for good?"~\footnote{For the readers convenience, we have replaced the original Berg's symbols by those used in the present paper.} 
Then he completely answered above questions deriving the desired trapping probability and the reaction rate at which a small diffusing particle is captured by a larger spherically symmetric absorber inside a spherical cavity (see details of his theory in Subsec.~\ref{subsec:Bergmodel}). 

However, "in genaral, macromolecules will not be reactive over their entire surfaces; instead only a very restricted region (the "active site")~\footnote{See also formal Definition~\ref{ActiveS} in Appendix.} will be directly involved in the reaction at issue."~\cite{Hippel85} Let us emphasize that these systems may be treated as a particular case of a vast class of physical an chemical systems encompass  the so-called {\it steric hindrance} of reactions, occurs when some parts ({\it chemically inert}) within a large molecule prevent or hinder chemical reactions that proceed easily only in a smaller ({\it chemically active}) parts. 

To make the task analytically tractable we will investigate a simple-minded model considering  small Brownian particles, diffusing towards a large enough spherical particle, which comprises an axially symmetric reactive site. It is significant that similar to the Berg case we suppose that this large particle is immobilized inside a spherical cavity. Note that, despite its simplicity, above physical model describes at least qualitatively, e.g., the association of a ligand to a reactive patch on a spherical molecule. This reactive patch could be the catalytic site on an enzyme or some other type of ligand-binding site on a spherical macromolecule (see detailed discussion in Ref.~\onlinecite{Berg85}).
We hope that this improved model can allow us to elucidate some facets on reaction-diffusion processes occurring inside biological cells.

In view of its importance, allow a rather lengthy quote from Eun's work.~\cite{Eun17}
"Here, we also note that to date, no exact solution has been found in a closed form of the Solc-Stockmayer model, and thus, most analytical theories employ some approximations,
except the one giving the exact solution as the infinite set of linear equations.~\cite{Traytak95} Among these analytical theories, we primarily use the analytical theory developed by Shoup, Lipari,
and Szabo (SLS), which is based on the constant-flux approximation.~\cite{Shoup81} In this paper, this theory is denoted the SLS theory.
The SLS solution is known to be one of the most accurate ones
with respect to the rate constant.~\cite{Traytak95}"

Below we will show that the dual series relations method used in Refs.~\onlinecite{Traytak94,Traytak95} can be  successfully applied to the problem considered in Ref.~\onlinecite{Eun17} too.

The main objective of the study is as follows: rigorous formulation and solution of the relevant mixed boundary value problem, which describes the diffusion-controlled reactions within the scope of Berg's model with a reactive particle having one axially symmetric reactive patch.

The paper is arranged as follows.
In the rest of introductory Sec.~\ref{sec:Intro} we present main aspects of classical Smoluchowski’s trapping theory. Section~\ref{sec:Prelim} contains
a summary on the subsequent generalizations of the Smoluchowski
theory to take into account geometrical confinement and chemical anisotropy effects.
Then, in next Secs.~\ref{sec:Statement}-\ref{sec:Dimensionless} we rigorously formulate the mixed boundary value problem for the steady-state diffusion equation to calculate the reaction rate within the scope of the one-patch Solc-Stockmayer model. Both geometrical and analytical components of the problem were considered in details.  Section~\ref{sec:LimCases} 
contains the treatment of two limiting cases of the problem to elucidate the way to attack a more complex original problem. General solution of the problem is found and discuss in Sec.~\ref{sec:GenSol}.
The exact semi-analytical solution to the posed mixed boundary value problem with the help of dual series relations method is given in Secs.~\ref{sec:RedDSR}-\ref{sec:SolISLAE}. Results of numerical calculations, including comparison to the obtained analytical approximations for the reaction rate, along with corresponding error estimates are presented in Sec.~\ref{sec:NumRes}.  Sec.~\ref{sec:methodOut} contains some discussion of the obtained results and a brief outline of the main ideas of the used approach.
We point out there the relationship with the generalized method of separation of variables, proposing a step by step algorithm of the solution procedure for the problem considered in this article. Finally, conclusions together with possible future extension of the present work are summarized in Sec.~\ref{sec:Conclusion}. In the Appendix we provide also some important physical assumptions~\ref{subsec:Backappendix} and  basic mathematical notations along with the background technical details of the used method~\ref{subsec:Basic}.

\subsection{Smoluchowski's trapping theory}\label{sec:Smol}

Since this article will develop Berg's model relies on the diffusion-controlled reactions theory which in its turn based on the standard 3D {\it Smoluchowski trapping model}, we briefly recall it.~\cite{Rice85} For the detailed discussion on the Smoluchowski approach and its physical background we refer the reader to a few comprehensive works on this subject.~\cite{Rice85,Szabo89,Ovchinnikov89}
Furthermore, to make this work maximally self-contained and facilitate understanding, we have also provided the assumptions underlying this theory in~\ref{subsec:Backappendix} of Appendix.

Let us consider the Brownian motion of small (point-like) guest particles
$B$ (from now on called $ B $-{\it particles} or $B$-{\it reactants}) towards the spherical surface of reactants $A$. Assume that the diffusion of $B$'s, describing by the free diffusion equation, takes place outside reactants $A$, while reaction occurs on the their surfaces only. Thus, we deal with typical reaction-diffusion process in a microheterogeneous system.~\cite{Seki01}

It is a matter of common knowledge that very often, in condense media (usually aqueous solutions in biology) reactions between reactants $A$ and $B$ can be treated as the 3D bulk  irreversible diffusion-controlled reactions described by the simplest reaction scheme~\cite{Rice85,Ovchinnikov89,Grebenkov23}
\begin{align}
	A+B\stackrel{k_f}{\longrightarrow }A + P\,, \label{Zv00}
\end{align}
where  $P$ designates some chemically inert reaction products and $k_f$ stands for the so-called {\it forward reaction rate constant}. One can see that reactions (\ref{Zv00}) are the {\it catalytic type reactions}~\cite{Oshanin05}, when activity of the reactants  $ A $ (catalyst) remains unchanged leading in turn to the decay of $B$-particles concentration. Further a particle $A$ it is also convenient to call a {\it reactive particle} (RP).~\cite{Traytak18}

Starting from the seminal Smoluchowski work the microscopic description of the diffusion-controlled reactions is based on some diffusion equation in terms of, e.g.,  trapping probability of $B$-particles $\rho\left( {\bf r}\right)$ (\ref{SurPr1}) posed in a 3D domain outside of the RP with
appropriate initial and boundary conditions (see ~\ref{subsec:Backappendix} of Appendix). Additionally, in this paper we suppose that {\it stabilization of the diffusion equation} (as time tends to infinity) holds and, therefore, the {\it steady-state approximation} is valid (see discussion of this question in Subsec.~\ref{subsec:AnalC}).
Explicit form of corresponding chemical kinetics equations are derived by means of microphysical modeling usually based on the Smoluchowski approach assuming that reaction surfaces of the RP are chemically isotropic. 

Originally the {\it Smoluchowski trapping theory} in its simplest form was proposed to describe coagulation of colloids.~\cite{Rice85,Doktorov19} Then it turned out that this theory has been successfully applied for studying most reactions in condensed phases.
Really, the theory is essential to elucidate a wide variety phenomena such as fluorescence quenching, radical-scavenger reactions, folding and stability of proteins, ligand binding to receptors, enzyme catalysis,~\footnote{It is evident, for instance, that the general scheme (\ref{Zv00}) becomes a well-known and widely used in biology the classical scheme for the irreversible {\it Michaelis-Menten kinetics} describing the  diffusion-controlled reactions between enzymes and substrates.~\cite{Dinner06,Bressloff14}}
drug delivery, etc.~\cite{Szabo89,Wiegel83,Lauffenburger93,Saltzman01,Saltzman04,Dinner06,Galanti16a} 

Once the trapping probability of $B$-particles $\rho\left( {\bf r}\right)$ is in hand, we can then calculate the microscopic reaction {\it trapping rate}  $k>0$, which is a number of $B$-reactants absorbing by the RP boundary or, generally speaking, its part per unit time.~\cite{Rice85} 
To solve the trapping problem uniquely Smoluchowski used {\it fully absorbing boundary condition} (\ref{Seki3}), supposing that $B$-particles are absorbed immediately once they reach the RP reactive boundary. Thus he found for the microscopic trapping rate~\cite{Rice85} 
\begin{align}
	k:=k_S=4\pi RD\,.  \label{Smol1}
\end{align}
Hereafter $R$ is the RP radius and $D$ is the translational diffusion coefficient of $B$-reactants. 

A crucial point is that within the scope of Smoluchowski's theory for the irreversible diffusion-controlled reactions (\ref{Zv00}) forward reaction rate constant $k_f$ has been identified with the microscopic trapping rate $k$ given by Eq. (\ref{Smol1}). However,  this is impossible to make in general case (see Remark~\ref{Rate} below at the end of Sec.~\ref{sec:rate}).

Despite the great success of the Smoluchowski theory to explain most aspects of the diffusion-controlled reactions there are many important questions still remain unsolved within its scope.

\subsection{Extensions of Smoluchowski’s theory} \label{subsec:Extensions}

It is now well-established by many authors that the validity of Smoluchowski’s model is restricted by a number of physically important reasons.
Particularly Smoluchowski’s theory is essentially a one-trap theory, which does not account for the {\it diffusive interaction} effects due to influence of neighboring particles including, fully absorbing, fully reflecting or partially reflecting entities. Interested readers can find details on the diffusive interaction concept, e.g., in  Refs.~\onlinecite{Traytak92,Traytak95c,Traytak08,Galanti16a,Grebenkov19,Piazza19,Traytak24}.

However, in the present paper we focus on the extra two essential limitations of the theory requiring refinements. Firstly the theory should be generalized for the case of reactions occurring in finite domains to include the diffusive interaction between the RPs and domains boundaries. Secondly it must be refined to describe the {\it chemical anisotropy} effects due to the sterically specific surface structure of the RPs, which can comprise active sites upon an otherwise inert surface.

Diffusion and then subsequent chemical reactions on the active sites belonging to the RPs surfaces
are crucial in the cell biology and drug delivery within a cell. Indeed "If the drug acts within the cell, it must move from the point of entry to the site of action. For some drugs, particularly for protein- and gene-based agents, this active site is within a specialized compartment or organelle in the cell; rates of transport within the cell are therefore important."~\cite{Saltzman01}

\section{Preliminaries} \label{sec:Prelim}

Before commencing the statement of the problem to be solved in the paper, a preliminary discussion of the published theoretical works concerning the subject at issue are perhaps helpful.

\subsection{The geometrical effects for reactions in a cavity} \label{subsec:cavities}

The investigations of diffusion transport of particles and diffusion-controlled reactions occurring  in cavities have a long and rich history. This is due to the fact that there are many important diffusion and reaction processes take place in natural and artificial bounded domains of various physico-chemical origin, which can be well modeled by cavities of various sizes and shapes. In particular, it is that quite often $B$-particles motion inside a biological cell can be characterized as normal Brownian diffusion with reasonable accuracy. Moreover, the diffusion-controlled reactions between $B$-particles and single or multiple enzymes that can be encapsulated inside a cell were of particular interest. So, following Ref.~\onlinecite{Sneppen05}, throughout this paper we use term "cavity" as a synonym to term "cell".

An enormous works on this topic has been published, and we refer the reader to extensive literature on these issues.
~\cite{Purcell77,Dubinko89,Ovchinnikov89,Bug92,Seki01,Tsao02,Price09,Schuss13,Bressloff14,Vazquez-Duhalt17,Grebenkov18,Kucherenko18,Holcman19,Berezhkovskii19,Zhang22,Bressloff22a,Bressloff22,Bressloff24,Piazza10,Traytak13,Poster13,Dzubiella13,Traytak15,Galanti16a,Galanti16b,Piazza19,Dzubiella20,Dolgushev21,Feng23,Dagdug24} 
However, it is significant that sometimes there is a misunderstanding of the crucial importance of the finite domain boundary influence during diffusion-controlled reactions. 
For example, in well-known textbook by Sneppen and Zocchi to describe the diffusion-controlled reactions occurring inside a spherical cell (see p. 182 of  Ref.~\onlinecite{Sneppen05}) the Smoluchowski trapping rate~\footnote{Eq. (\ref{Smol1}) was derived for an isolated ideal absorber in a free physical space} used directly without correction factor. 
We have shown that the above result does not valid for the case at issue~\cite{Poster13,Traytak15,Grebenkov19,Piazza19} and our conclusion was supported also in Ref.~\onlinecite{Vazquez-Duhalt17}:"...the Smoluchowski reaction rate constant does not consider the nanoreactor geometry and the crowding effects."

\subsection{The chemical anisotropy effects} \label{subsec:CAnieffects}

Theoretical works, concerning trapping reactions on chemically anisotropic reactants have also been a subject of large interest in many years (for details the reader is referred to, e.g., Refs.~\onlinecite{Eun17,Popescu19,Doktorov19,Doktorov23,Dagdug24} and references therein).

The research of the chemical anisotropy effects stems from the seminal work by Alberty and Hammes Ref.~\onlinecite{Alberty58} devoted
to application of the theory of diffusion-controlled reactions to enzyme kinetics. Since then, this issue has been studied by many authors. 

Developing the Smoluchowski trapping theory, Hill derived the diffusion flux of $B$-particles with  isotropic reactivity absorbing by a circular active site of radius $a$ located on an otherwise inert plane.~\cite{Hill75}
\begin{align}
	k_H = 4 \pi a D \,. \label{Hill75}
\end{align}
It turned out that the for large RP radius Hill's formula (\ref{Hill75}) is a good approximation, therefore it has been widely used in biological applications, e.g., in the Berg-Purcell theory describing ligand-binding to the surface receptors on a spherical cell.~\cite{Purcell77,Wiegel83,Hippel85,Berg85,Berg93}
We would like to draw attention to a few interesting theoretical studies that generalize Hill's result.
~\cite{Strieder08,Strieder09,Strieder12,Eun20b} Note, however, that some authors consider the  {\it model of Schmitz-Schurr} (or {\it model of reactive hemisphere})~\cite{Schurr76} more realistic than the Hill circular active site model.~\cite{Shoup81,Shushin00,Barzykin01,Barzykin01b}

Subsequently, with the help of different approaches, the theory was generalized to the cases of two~\cite{Pritchin85,Barzykin07,Lee09,Shoup14,Doktorov23} and many~\cite{Seki01,Gopich13,Berg16,Eun18,Bernoff25,Grebenkov25b} active sites located on an inert sphere.

This is a rather difficult mathematical problem and often the influence of diffusion on the kinetics of ligand binding to a macromolecule with two active sites is considered for a simple
model where, in the reaction-controlled limit, there is no cooperativity and hence the sites are independent~\cite{Gopich19} or using the so-called {\it SLS approximation} (or {\it constant flux approximation}).~\cite{Shoup81,Shoup14} 

In his review entitled in a rather catchy manner "An unprecedented revolution in medicinal chemistry driven by the progress of biological science" Chou paid great attention to the reaction occurring on reagents with anisotropic reactivity.~\cite{Chou17}  Note that he provided a detailed list of references to his early theoretical works.

Mention should also be made on significant efforts to solve problems under consideration by numerical and simulation methods.~\cite{Luty91,Luty92,McCammon05,McCammon12,Mascagni01,Eun17,Eun20}
We would like to draw special attention among these studies to Ref.~\onlinecite{Eun17}, where the {\it finite element method} was employed to attack problems closely related to those treated in the present research.

Finally, an additional point to emphasize is that for simplicity sake we do not consider the complications caused by the fact that $B$-particles can also have anisotropic reactivity. However, the reader for completeness can find some obtained results regarding this generalization, e.g., in Refs.~\onlinecite{Stockmayer73,Doi75,Temkin84,Karplus87,Zhou97,Barzykin01,Barzykin01b,Seki01,Doktorov22}.

In 2010 Vazquez considered the diffusion-controlled reactions of {\it annihilation type} when diffusing substrate $S$ is irreversibly transformed into the inert product catalyzed by the immobile enzyme $E$:~\cite{Vazquez10}
\begin{align}
	E+S\stackrel{k}{\rightarrow} P\,. \label{Lap1a}
\end{align} 
To account crowding effects and enzyme chemical anisotropy due to active sites he supposed that the reaction rate (\ref{Smol1}) should be modified to the following form
\begin{align}
	k = 4\pi R_{eff} D \,, \label{Vazquez}
\end{align}   
where  $R_{eff}$ is some the {\it effective size} of the enzyme active site. Then a heuristic approach was suggested to find value of  $R_{eff}$. 
The calculation of the effective size $R_{eff}$ for the axially symmetric one active patch {\it Solc-Stockmayer model} can be readily reduced to that for the trapping rate (see Eq. (\ref{EffR2}) in Subsection~\ref{subsec:DimLrate}).
Vazquez's paper was considerably generalized in our subsequent paper devoted to diffusion-influenced reactions in a hollow nano-reactor with a circular axially symmetric hole.~\cite{Traytak15}

Also, the so-called {\it narrow escape problems} and {\it target search problems} are worth mentioning especially as well-developed and yet a very close in mathematical technique field of research (see, e.g., Refs.~\onlinecite{Holcman06a,Schuss12,Bressloff14,Schuss15,Greengard20,Oshanin24} and literature therein).
Nevertheless, it should be highlighted the significant difference between the narrow escape problem and the problem under consideration. Really for the narrow escape problems appropriate mixed boundary conditions are imposed upon the cavity wall, but not on the RP surfaces.

To solve the mixed diffusion problems the dual series relations method first was proposed in our papers Refs.~\onlinecite{Traytak94,Traytak95} and then has been used successfully to calculation appropriate trapping rates in a number of subsequent works on different axially symmetric mixed boundary value reaction-diffusion problems.   ~\cite{Tachiya95a,Tachiya95b,Barzykin01,Holcman06a,Barzykin07,Price07,Lee09,Traytak15,Berezhkovskii17,Popescu19} 

Thus, if to summarize everything above we conclude, the modified theory of the diffusion-controlled reactions should bring two major modifications with respect to the classical Smoluchowski theory. First, it has to deal with diffusion-controlled reactions occurring in a 3D bounded, rather than infinite, domain. Second, it should takes into account the chemical anisotropy of the RP surface.

\section{Mathematical statement of the problem}\label{sec:Statement}

From the beginning it should be stressed that we tried to pursue our mathematical statement of the problem to be close in notations and definitions to that given in prior works.~\cite{Masagni04,Eun17,Grebenkov19,Eun20} Moreover, as is customary in mathematical physics, we will formulate the reaction-diffusion problem, separating its geometrical and analytical components.

\subsection{Geometrical components} \label{subsec:GeomCom}

Clearly, diffusion and the chemical kinetics of $B$-particles depend drastically on the geometry of the given host medium, so first the reaction-diffusion problem at issue is supposed to satisfy some geometric assumptions (see main physical assumptions in Appendix~\ref{subsec:Backappendix}).

\begin{figure}[h!]
	\centering
	\includegraphics[width=0.30\textwidth]{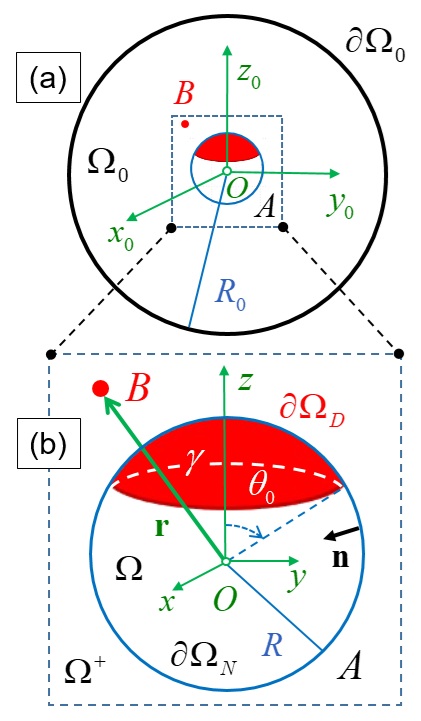}
	\caption{Geometric sketch of the axially symmetric one-active patch Solc-Stockmayer model. Panel (a): A test $B$-reactant and their source located at the outer boundary $\partial \Omega_0$  are given in black. Panel (b): The RP comprises  an inert sphere (blue) and an active site, which is modeled by a spherical cap with opening angle $\theta_0$ (red). The inner boundary $\partial \Omega$ (blue) 2-partition: $ \left\{ \partial \Omega_{D}, \partial \Omega_{N}\right\} $. In the both panels Cartesian coordinates are depicted in green.}
	\label{fig1}
\end{figure}

Consider a spherical cavity $\Omega_0 \subset \Bbb{R}^3$ of radius $R_0$, which center is located at a fixed point $ O $.
It is expedient to use a {\it global Cartesian coordinate system} $\{ O; x_0, y_0, z_0\} $ with the origin $O=\mathbf{0}= \left(0, 0, 0 \right)$ attached to the spherical cavity center (see Fig.~\ref{fig1} (a)). Denote the radius-vector corresponding to a current point $P$, describing the location of a $B$-particle, by $ \mathbf{r} $. At that its distance from the origin is $  r = \Vert \mathbf{r} \Vert $, where $\Vert \cdot \Vert $ stands for the common Euclidean norm in $ \Bbb{R}^3 $. Clearly, in the chosen coordinates the spherical cavity domain $\Omega_0$ with the boundary $\partial \Omega_0$ are described analytically as follows: $\Omega_0= \lbrace  \mathbf{r} \in {\mathbb{R}}^3: \Vert \mathbf{r}  \Vert < R_0 \rbrace$ and $\partial \Omega_0 = \lbrace  \mathbf{r} \in {\mathbb{R}}^3: \Vert \mathbf{r}  \Vert = R_0 \rbrace$. 
In other words we will consider physical-chemical processes that takes place inside an open 3-ball  $B^3=\Omega_0$ of radius $R_0$ bounded by the 2-sphere $S^2=\partial \Omega_0$.
The main cross-section of the ball $B^3$ at issue is depicted in  Fig.~\ref{fig1} (a).

In turn for the sake of simplicity we consider a concentric spheres geometry, depicted in Fig.~\ref{fig1}. Moreover, assume that an immobile spherical RP (also often called {\it target sphere}~\cite{Lindsay17}) of the reaction radius $R< R_0$  placed into the same origin $O$ (see Fig.~\ref{fig1} (a)). This allows us to identify the RP {\it local Cartesian coordinates}  $\{ O; x, y, z\} $ with above global system. The RP - $\Omega$ is entirely englobed by the cavity domain 
$\Omega_0$: $ \overline{\Omega} \subset \Omega_0 $, 
i.e., $\Omega = \{ {\bf r}\in \Bbb{R}^3~:~ \| {\bf r}\| < R < R_0\}$.

Using Fig.~\ref{fig1} it is evident that the spherical shell domain $\Omega^+$ and the exterior of the RP $ \Omega^-$ can be given as the complements of the RP domain $\overline{\Omega }$ in sphere $\Omega_0$ and $\R^3$, respectively:
\begin{align}
	\Omega^+:= \Omega_0\backslash  \overline{\Omega }\,, \quad  \Omega^-:= \R^3\backslash \overline{\Omega }\,. \label{Dom1}
\end{align}
Explicitly for these domains one has: $ \Omega^+= \{ {\bf r}\in \Bbb{R}^3~:~R < \| {\bf r}\| < R_0\}$ and $ \Omega^-= \{ {\bf r}\in \Bbb{R}^3~:~ \| {\bf r}\| > R\}$. Besides, the shell domain $\Omega^+$ may be expressed as
\begin{align}
	\Omega^+= \Omega_0\cap \Omega^- \subset  \Omega^- \,. \label{Dom2}
\end{align}

Before proceeding further recall the following 
\begin{definition}\label{defin1}
	For structureless point-like $ B $-particles the {\it configuration manifold} is a 3D domain composed of all $ B $-particle positions $ \mathbf{r} $.~\cite{Traytak24}
\end{definition}

Since point-like $B$-particles freely diffuse in the spherical shell domain $\Omega^{+}$, we can call it the {\it configuration manifold}.~\cite{Traytak24} Evidently the configuration manifold is bounded by the $ 2 $-connected boundary (see Fig.~\ref{fig1} (a)) including the {\it outer boundary} $\partial\Omega_0$ and the {\it inner boundary} $\partial\Omega$ (RP surface) that is 
\begin{align}
	{\partial \Omega^{+}}=\partial{\Omega_0}\cup\partial{\Omega }\,.
	\label{Bound1a}
\end{align}   

To study problems for the concentric spheres geometry it is convenient to use the spherical polar coordinate system $ \left\{O; r, \theta ,\phi \right\} $ associated with the chosen above Cartesian coordinates (see Appendix~\ref{subsec:Basic}).

Assume that the RP $A$ contains one active spherical-cap patch absorbing $B$'s with {\it opening angle} ({\it angular size}) $\theta_0$, such that the physical system "RP-cavity" exhibits axial symmetry around the $z$-axis (see Fig.~\ref{fig1}).
\begin{remark} \label{RedC}
	Nowadays the active patches most commonly are depicted in red~\cite{Eun18,Traytak18,Alejo22,Lawley23,Dagdug24} so in this study we adopt red color for the absorbing patch on the RP surface.
\end{remark} 

In the present case  the union (\ref{Part1}) constitutes the  $2$-partition of boundary comprises the pair $ \left\{ \partial \Omega_{D}, \partial \Omega_{N}\right\} $. 
Here and after we use the subscripts "$D$" and "$N$"  to indicate the parts of the boundary where Dirichlet (\ref{Seki3})  and Neumann (\ref{Seki4}) boundary conditions are imposed, respectively.  

Thus, we assume that the reaction surface consists of two capped parts: (i) absorbing $\partial \Omega _D\subset
\partial \Omega $ and (ii) reflecting $\partial \Omega _N=\partial \Omega\backslash
\partial \overline{\Omega }_D$ (see Fig.~\ref{fig1} (b)). In the above polar spherical coordinates (\ref{SphC1}) let us denote the angular size of the active site by $\theta_0$. Then, clearly, we have explicit
\begin{eqnarray}
	\partial \Omega _D=\left\{ r=R, \,0 < \theta < \theta _0, \, 0 < \phi <2\pi
	\right\} \,,  \label{Seki1a}
	\\	
	\partial \Omega _N=\left\{ r=R,\,\theta _0<\theta <\pi, \, 0 < \phi <2\pi
	\right\}\, .  \label{Seki1b}  
\end{eqnarray}

There is another important geometrical component. Consider a circular cone surface (it makes an angle of $ \theta_0 $ with its symmetry $ z $-axis with the vertex at the origin $ O $) cuts the circle $\gamma = \lbrace r = R \rbrace \times \lbrace  \theta = \theta_0 \rbrace$ on the RP surface (see Fig.~\ref{fig1} (b)).
One can see that this circle is nothing but the edge between two parts of the RP boundary:  $\gamma= \overline{\Omega }_D \cap \overline{\Omega }_N$. Edge $\gamma$ plays an important role for the uniqueness of the classical solution (see next subsection).

\subsection{Analytical components}  \label{subsec:AnalC}

As is shown in Ref.~\onlinecite{Traytak94} the chemical asymmetry of the RP leads to a decrease of the transition period for the reaction rate.
Therefore, further we will consider steady-state diffusion-controlled reactions only. Moreover, it should be pointed out two important consequences of this result.
First it has been proven that, the diffusive interaction is the most profound in the steady-state regime that allows us to estimate the maximum affects to the reaction rate.~\cite{Traytak24} Secondly, when studying the diffusive interaction, treatment of the steady state allows us to simplify the problem essentially.

The steady-state {\it local concentration} (or {\it number density}) $ n_B(\bf r) $ of point-like $B$-reactants at location $ \bf r $ of the configuration manifold $\Omega ^{+}$ is usually used to investigate the steady-state microscopic diffusion-reaction model.~\cite{Rice85}
However, often in applications function $ n_B \left( {\bf r}\right)$ is scaled with its bulk concentration $ c_B $ prescribed at the finite or infinite~\cite{Shushin00,Barzykin01,Barzykin01b} outer boundary $ \partial \Omega_0 $ 
\begin{align}
	\rho \left( {\bf r}\right) := n_B \left( {\bf r}\right)/c_B\,, \quad c_B:=\lim_{{\bf r} \to \partial \Omega_0-}n_B \left( {\bf r}\right)\,. \label{SurPr1} 
\end{align}
The above scalar field $\rho: \Omega^{+} \to \left(0, 1\right) $ has simple physical meaning of the  {\it survival probability}, i.e., the probability of finding a $B$-particle at position ${\bf r} \in \Omega^{+}$.~\cite{Rice85,Piazza19} 

Note here that Berg used this very function for describing diffusion to capture in his model.~\cite{Berg93} Moreover, it is important to stress that, taking into account finiteness of the configuration manifold under study $\Omega ^{+}$ (\ref{Dom1}), the probabilistic interpretation should be used for functions, describing reaction-diffusion behavior of $B$-particles appropriately.

One can see  that because of axially symmetric geometry of the problem function $\rho \left( {\bf r}\right)$ is also axially symmetric and it does not depend on the azimuthal angle $\phi$, which will be omitted in the remaining text.

Therefore, we can write down the steady-state governing diffusive system:~\cite{Barzykin01} continuity equation in the absence of sources (\ref{Lap0a}) and the classical constitutive relation (\ref{Lap1a}) (the first Fick’s law of diffusion) defined in the configuration manifold $\Omega^{+}$ as
\begin{eqnarray}
	\bm \nabla \cdot \mathbf{j} =0\,, \label{Lap0a} \\ 
	\mathbf{j} = -D{\bm \nabla} \rho  \,. \label{Lap1a} 
\end{eqnarray}
Hereafter
$ \mathbf{j}\left( {\bf r}\right) $ is the local diffusive flux at point $ \mathbf{r} $ and $ \bm{\nabla} $ denotes the gradient operator in $\mathbb{R}^3$.

The Fick steady-state diffusive system Eqs. (\ref{Lap0a}), (\ref{Lap1a}) immediately leads to the Laplace equation
\begin{align}
	-D{\bm \nabla}^2 \rho =0 \quad \mbox{in} \quad \Omega ^{+} \,.	\label{Lap1}
\end{align}
Hereinafter  $ {\bm\nabla} ^2:= {\bm\nabla} \cdot {\bm\nabla} $ is the scalar Laplace operator.

For the problem under consideration we suppose that the bulk value of concentration $c_B$ (\ref{SurPr1})  is maintained over the outer boundary
\begin{align}
	\left. \rho \left( {\bf r}\right) \right| _{\partial \Omega_0 }= 1 \,,  \label{Seki5}
\end{align}

In Eq. (\ref{Seki5}) the boundary function has the same values along the whole outer boundary, that is we deal with the {\it Dirichlet boundary condition}.
Physically (\ref{Seki5}) means that the source of $B$-reactants is located on the cavity boundary $\partial \Omega_0$ (see Fig.~\ref{fig1}).

Chemical anisotropy of the RP surface can be modeled by the inhomogeneous boundary condition. 
As it was assumed above the reaction is diffusion-controlled, so on the active site $\partial \Omega_D$ we can prescribe the  {\it fully absorbing} boundary condition~\cite{Rice85,Traytak24}
\begin{align}
	\left. \rho\left( {\bf r}\right) \right| _{\partial \Omega _D}=0\,.  \label{Seki3}
\end{align}

Clearly, on the inert part $\partial \Omega _N$ the {\it fully reflecting} boundary condition should be imposed
\begin{align}
	\left. \left( {\mathbf n} \cdot {\mathbf j} \right) \right|
	_{\partial \Omega _N}=0\,.  \label{Seki4}
\end{align}
In mathematics above conditions are referred to as {\it homogeneous Dirichlet} (\ref{Seki3}) and {\it homogeneous Neumann} (\ref{Seki4}) boundary conditions. 

Thus, in accordance with Definition~\ref{defin2} (see Appendix~\ref{subsec:MixedBVP}), the posed diffusion problem (\ref{Lap1})-(\ref{Seki4}) is the interior Dirichlet-Neumann  {\it proper mixed boundary value problem}.~\footnote{A proper mixed Dirichlet–Neumann boundary-value problems for the Laplace equation is also called {\it Zaremba problem}.~\cite{Sneddon66}} Note that, henceforth in the paper word "proper" will be dropped since we treat only proper mixed boundary value problems.

It should be emphasized that by solution to the posed mixed boundary value problem we understand only its classical solution. Concerning  the existence and uniqueness of the classical solution to the mixed boundary value problems we refer the interested reader to, e.g., Ref.~\onlinecite{Volpert85}.
And in the end of the problem formulation, we should dwell on a one more important point. Let us note a significant and yet subtlety of mixed boundary value problems that is not paid attention to by researchers studying the theory of diffusion-controlled reactions. In literature, this subtle mathematical point  is referred to as “edge conditions” at the edge point $\theta=\theta_0$ (see brief discussion in subsection~\ref{subsec:profile}).
These conditions are necessary to guarantee the uniqueness of the solution and  good convergent numerical procedures 
because  there exists a geometric singularity at the edge points. Fortunately for the problem under consideration is caused by integrable essential discontinuity of the normal derivative on the edge.

\section{The microscopic trapping rate}\label{sec:rate}

A central quantity of interest from the solution of the posed problem is the total flux through the RP surface, defined as

According to the Smoluchowski theory to describe the kinetics of the irreversible diffusion-controlled reactions (\ref{Zv00}) one should estimate the microscopic reaction rate $k$. With the help of known solution $ \rho \left( {\bf r}\right) $, the  microscopic trapping reaction rate can be calculated straightforwardly by the general formula~\cite{Rice85} 
\begin{align}
	k  =\oint\limits_{\partial \Omega}\left. \left( {\mathbf n} \cdot \mathbf{j} \right) \right| _{\partial\Omega}dS \,.  \label{Lap2}
\end{align}
Hereinafter $ {\mathbf n} \left( {\bf r}\right) $ being the normal unit vector pointing outward of $ \Omega ^{+} $ at its spatial point of the boundary $ {\mathbf r} \in \partial\Omega $ (see Fig.~\ref{fig1} (b)) and $d{S}$ is differential element of the boundary $\partial\Omega$ area. 

\begin{remark} \label{Form}
	One can see that for the above Dirichlet-Neumann mixed problem formula (\ref{Lap2}) may be directly simplified to 
	\begin{align}
		k = \int\limits_{\partial \Omega_{D}}\left. \left( {\mathbf n} \cdot \mathbf{j} \right) \right| _{\partial
			\Omega_{D}}dS \,.  \label{Seki6}
	\end{align}
\end{remark}

Below in Secs.~\ref{sec:RedDSR}-\ref{sec:SolISLAE} we will present a powerful mathematical approach to calculate the trapping rate (\ref{Seki6}) for the assumed geometry (\ref{Seki1a}), (\ref{Seki1b}) of the problem under consideration. 

\begin{remark} \label{Rate}
	It is significant that in general case the microscopic trapping rate (\ref{Lap2}) does not coincide with the reaction rate constant $ k_f $ defined by Eq. (\ref{Zv00}), which is a fundamentally macroscopic value.~\cite{Traytak92,Traytak24} 
	So, investigating diffusive interaction effects, generally speaking,  we cannot use the term "rate constant" for $ k $ given by Eq. (\ref{Lap2}).
\end{remark}

\section{Dimensionless form of the problem} \label{sec:Dimensionless}

We have completely formulated above the Dirichlet-Neumann mixed problem (\ref{Lap1})-(\ref{Seki5}) for the dimensional one-patch Solc-Stockmayer model. However, to carry out further mathematical analysis it is convenient to recast this problem and, thereby, the associated microscopic trapping rate (\ref{Seki6}) in a dimensionless form using the chosen spherical coordinate system (\ref{SphC1}).

\subsection{The reaction-diffusion problem}\label{subsec:ReacDiff}

However, below, we are going to use another function namely {\it trapping probability}, which seems to be more appropriate to describe diffusive interaction effects~\cite{Traytak24}
\begin{align}
	u: \Omega^+ \to \left(0, 1\right)\,, \quad \mbox{where} \quad	u \left( {\bf r}\right) = 1- \rho\left( {\bf r}\right)\,.  \label{Zv02a}	
\end{align}
It is clear that  $u\left( {\bf r}\right)$ is the probability that a reaction occurs when any $B$ particle undergoes a contact with the active site of the RP. Simply speaking, a trapping probability, is equal to a  probability of $B$-particles collision with the active site.

Nevertheless, following tradition,~\cite{Rice85} further we will call function (\ref{Zv02a}) local concentration if there is no confusion to be appeared.

Let us introduce now the dimensionless radial coordinate $ {\bf r} \mapsto \bm \xi := (\xi, \theta)  $ by rescaling of the independent spatial variable. Thus for the dimensionless spacial variable $\xi$ and corresponding domains and their boundaries defined in Subsection~\ref{subsec:GeomCom} one obviously has 
\begin{eqnarray}
	\xi =r/R\,, \quad \xi \in \left(1,1/\epsilon\right) \,,	\label{Resc1} \\
	\Omega \to \Omega_{\xi}= \{ 0<\| {\bm \xi }\| < 1\}\,, \label{Resc2a}\\
	\Omega_{0} \to \Omega_{\xi}^{0}= \lbrace  0< \Vert \bm \xi   \Vert < 1/\epsilon \rbrace \,,	 \label{Resc2}\\
	\Omega^{-} \to \Omega_{\xi}^{-}=\lbrace  \Vert \bm \xi   \Vert > 1 \rbrace\,,   \label{Resc3}\\
	\Omega^{+} \to \Omega_{\xi}^{+}=\lbrace 1<  \Vert \bm \xi   \Vert < 1/\epsilon \rbrace\,, \label{Resc4}\\
	\partial\Omega_{\xi}^{+}=\partial\Omega_{\xi} \cup \partial\Omega_{\xi}^{0}\,, \label{Bound1}\\
	\partial\Omega_{\xi}= \lbrace  \Vert \bm \xi   \Vert = 1\rbrace\,, \quad	\partial\Omega_{\xi}^{0}= \lbrace  \Vert \bm \xi   \Vert = 1/\epsilon \rbrace  \,. \label{Bound2}
\end{eqnarray}
Let us focus that hereinafter, to make dimensionless parameters {\it thickness ratio} $\epsilon$ and the {\it shell thickness} $h$ be less than unity, we normalized  the RP size $R$ and thickness $R_0-R$ by $R_0$: 
\begin{eqnarray}
	0<\epsilon := {R}/{R_0} < 1\,, \label{thickness2}\\
	0<h:= \left(R_0-R \right)/R_0= 1- \epsilon <1 \,. 	\label{thickness1}		  
\end{eqnarray}

Whereas the local concentration (\ref{Zv02a}) and Laplace's equation (\ref{Lap1}) take the dimensionless form
\begin{eqnarray}	 
	u\left(\bm \xi \right) =1-\rho\left(\bm \xi \right)\,, \quad u\left(\bm \xi \right) \in \left( 0,1\right)\,,   \label{LE1a}\\
	- {\bm\nabla}_{\bm \xi}^2u=0 \quad \mbox{in} \quad \Omega_{\xi}^{+}\,. \quad \label{LE1}
\end{eqnarray}
The Laplacian in Eq. (\ref{LE1}) reads 
\begin{eqnarray}
	{\bm\nabla}_{\bm \xi}^2:={\nabla}^2_\xi + \frac{1}{\xi^2}{\nabla}^2_{\theta} \,, \qquad \quad \nonumber \\	
	\mbox{where} \quad	{\nabla}^2_\xi:=\frac{1}{\xi^2}\frac{\partial}{\partial \xi}\xi^2 \frac{\partial}{\partial \xi}\,, \quad {\nabla}^2_{\theta}:=\frac 1{\sin \theta }\frac \partial {\partial \theta }\left(
	\sin \theta \frac{\partial }{\partial \theta }\right)   \qquad \nonumber
\end{eqnarray}
are the spherically symmetric Laplacian and the axial symmetric Laplace-Beltrami operator, respectively. 

Therefore the boundary conditions may be rewritten as follows:
\begin{eqnarray}
	\left. u\right| _{\xi \to 1/\epsilon -} \to 0\, \quad \mbox{for} \quad  0\leq
	\theta \leq \pi\,,    \label{LE2}\\
	\left. u\right| _{\xi =1+}=1\,, \quad 0\leq \theta <\theta _0 \,, \label{MBC1}\\ 
	\left. \frac{\partial u}{\partial \xi}\right| _{\xi =1+}=0\,,  \quad \theta _0<\theta \leq \pi\, .  \label{MBC2}	
\end{eqnarray}
In this way, we have recast the original Dirichlet-Neumann mixed problem (\ref{Lap1})-(\ref{Seki4}) to its dimensionless form (\ref{LE1})-(\ref{MBC2}).

\subsection{Representation of the microscopic trapping rate}\label{subsec:DimLrate}

Then, recasting formula (\ref{Lap2}) with respect to $ u \left({\bm \xi} \right) $ in an exterior neighborhood of the RP surface, one can easily obtain the desired microscopic trapping rate by the integral over the unit sphere 
\begin{align}
	k \left( \theta_0; \epsilon\right) = - 2\pi 	\int_{0}^{\pi} \left. \frac{\partial u}{\partial \xi}\right| _{\xi=1+} \sin\theta d\theta\,.	\label{dm1n}
\end{align}

It is obvious from a physical view point that for small shell thickness $h$ (\ref{thickness1}) diffusive interaction effects become important. So one should introduce an appropriate correction factor to the Smoluchowski absorbsion rate (\ref{Smol1}).
Therefore, the explicit form of the reaction rate (\ref{dm1n}) it is convenient to seek as an ansatz
\begin{align}
	k = k_{S} J\left( \theta_0; \epsilon\right)\,,	\label{dm1na}
\end{align}
where the quantity $ J\left( \theta_0; \epsilon\right)$ is commonly called the {\it rate correction factor} (RCF)~\cite{Traytak24} (or {\it screening coefficient}).
\begin{remark} \label{Sherwood}
	Note in passing that the RCF is a dimensionless parameter, which may be treated also as the corresponding Sherwood number for the RP.~\cite{Traytak92,Borovkov24}
\end{remark}

Expression (\ref{dm1na}) generalizes Eq. (\ref{Smol1}) and for all $\left( \theta_0, \epsilon\right) \in \left\{ 0\leq \theta _0 \leq \pi \right\}\times \left\{ 0 < \epsilon < 1 \right\}$ the following relations hold true 
\begin{eqnarray}
	0 < J\left( \theta_0; \epsilon\right)  <1  \,, \quad \label{CorC1}\\ 
	J\left( \theta_0; \epsilon\right)  \to 1 \quad \mbox{as} \quad \left( \theta_0, \epsilon\right) \to \left(\pi, 0\right) \,. 	\label{CorC2}  
\end{eqnarray}
Double limit (\ref{CorC2}) simply means that the natural requirements must be held: for a given size of the RP $R$ isotropic reactivity when $\theta_0 =\pi$ and diffusive interaction  disappears at large $R_0$.

So, it turns out that for our case the effective size defined by Eq. (\ref{Vazquez}) can be written as
\begin{align}
	R_{eff} = R J\left( \theta_0; \epsilon\right)\,.\label{EffR2}
\end{align}
Below we will focus our efforts at the explicit calculation of the RCF $ J\left( \theta_0; \epsilon\right)$ and, in this way, effective radius (\ref{EffR2}).

It has been known that posed problem is impossible to solve with a standard method of separation of variables. So for its solution we will involve a modification of the so-called generalized method of separation of variables (see discussion in Sec.~\ref{sec:methodOut}).

\begin{remark} \label{synonyms}
	And the last, but not least, due to simple proportionality relationship (\ref{dm1na}) between the reaction rate $ k\left( \theta_0; \epsilon\right)$ and the rate correction factor $ J\left( \theta_0; \epsilon\right)$, further these two concepts can be used as synonyms.
\end{remark}

\section{Limiting cases of the problem} \label{sec:LimCases}

Before proceeding to the study of the above posed general problem let us briefly dwell on two simpler particular cases of the Dirichlet-Neumann mixed problem (\ref{LE1})-(\ref{MBC2}) to elucidate the method of its solution used further.

It is pertinent to quote here Ref.~\onlinecite{Kovalenko86}
"The main peculiarity of the problems under mixed boundary conditions is as follows: one should first learn how to deal with the relevant nonmixed problems, and only then to a solve their mixed counterparts." Therefore, to elucidate the possible pitfalls arising when solving the posed mixed problem, consider here the relevant so-called {\it toy model}. By this we mean a greatly simplified model, which helps us for approaching more complex original problem. Evidently, Berg's model plays  role of the toy model in our research.

\subsection{The spherically symmetric Berg's model}\label{subsec:Bergmodel}

\begin{figure}[h!]
	\centering
	\includegraphics[width=0.24\textwidth]{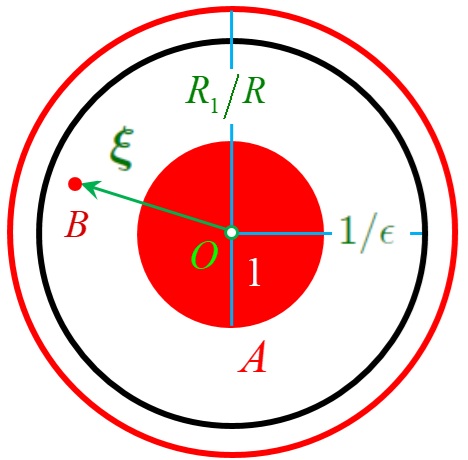}
	\caption{Geometry of the spherically symmetric Berg's model~\cite{Berg93}	for the limiting case when the RP (red) is chemically isotropic and ideally absorbing. The spherical shell adsorber is also shown in red. Hereinafter $\epsilon:=R/R_0$ (\ref{thickness2}).}	
	\label{toy}
\end{figure}

For descriptive reasons, following Berg's book,~\cite{Berg93} let us consider the spherically symmetric model of diffusion-controlled reactions occurring on the ideally absorbing chemically isotropic ($\theta_0 = \pi$) RP inside a spherical cavity.  Geometry of the Berg model is depicted in Fig.~\ref{toy} and analytically described as follows: (1) the RP with surface $\partial\Omega_{\xi}$ ($\Vert \bm \xi \Vert = 1$); (2) cavity wall $\partial\Omega_{\xi}^{0}$ ($\Vert \bm \xi \Vert = 1/\epsilon$)  and, (3)  a spherical shell adsorber of radius $\Vert \bm \xi \Vert =R_1/R$. 

To avoid unduly complication the problem, below we will assume that $R_1/R \to \infty$ and,
therefore, consider the behavior of the local concentration (\ref{LE1a}) $u\left(\bm \xi \right)$  in the spherical layer $1<\xi<1/\epsilon$.


So axially symmetric mixed Dirichlet-Neumann problem (\ref{LE1})-(\ref{MBC2}) for geometry given in Fig.~\ref{toy} is simplified to the spherically symmetric Dirichlet problem of the form
\begin{eqnarray}
	-{\nabla}^2_\xi u=0 \,, \label{SSE1}\\ 
	\left. u\right| _{\xi \to 1/\epsilon-} \to 0\,, \quad \left. u\right| _{\xi =1+}=1\, . \label{SSE2}
\end{eqnarray}
Dirichlet problem (\ref{SSE1}), (\ref{SSE2}) has the exact solution
\begin{align}
	u\left(\xi\right) = \frac{\epsilon}{1-\epsilon}\left( \frac{1}{\epsilon\xi} - 1\right)  \,. \label{SSE3}
\end{align}

From this it obviously results that for the steady state all $B$-particles penetrated into the cavity at the outer boundary $ \left\{ \xi = 1/\epsilon \right\}$ will be absorbed by the RP on its surface $ \left\{ \xi = 1 \right\}$.

It is notable that although $1 \le \xi < + \infty$ and $0< \epsilon < 1$  but, taking into consideration the range of function $u\left(\xi\right)$ (\ref{SSE3}) the following inequality $0< \epsilon\xi <1$ must be hold. Thus for the RCF (\ref{dm1na}) we easily obtain~\cite{Berg93,Piazza19}
\begin{align}
	J\left( \pi, \epsilon \right)= \frac{1}{1-\epsilon} = \frac{1}{h}\,. \label{SSE4}
\end{align}

Here we also expressed the RCF with the aid of spherical shell domain $\Omega_{\xi}^+$ thickness  $h$ (\ref{thickness1}).
So the rate  $k: \left(0, 1\right) \to \left(1, +\infty\right)$ is the monotonically increasing function such that
$$
k \left( \epsilon\right) = k_S\frac{1}{1-\epsilon}> k_S= \inf_{\epsilon \in \left(0, 1\right)}k \left( \epsilon\right) \,. 
$$
It is evident that in case of unbounded configuration manifold $\Omega ^{-}$ (as $\epsilon \to 0$) Eq. (\ref{dm1na}) for the reaction rate leads to the steady-state Smoluchowski rate constant $k_S$.

\begin{remark} \label{ToyMod}
	On the other hand the RCF (\ref{SSE4}) becomes very large as the RP approaches the inner surface of the cavity. Therefore, it follows from considered toy model that we can face rather subtle mathematical difficulties when $\epsilon \to 1$ (or $h \to 0$).
\end{remark}

A mathematical description of the problem for thin enough shell spherical domain ($h \ll 1$) may be performed by means of perturbation theory, which is, however, beyond the scope of the present paper.

\subsection{Unbounded configuration manifold}

According to relations (\ref{thickness1}), (\ref{thickness2}) and definitions (\ref{Resc3}), (\ref{Resc4}) it is obvious that $\Omega_{\xi}^{+} \to \Omega_{\xi}^{-}$ as $\epsilon \to 0$.

If the RP reactivity depicted in Fig.~\ref{fig1} (b), the internal mixed problem (\ref{LE1})-(\ref{MBC2}) is reduced to the associated external mixed problem 
\begin{eqnarray}
	-{\bm\nabla}_{\bm \xi}^2 u=0 \quad
	\mbox{in} \quad \Omega_{\xi}^{-}\,,  \quad \label{TD1} \\
	\left. u\right| _{\xi =1+}=1\,, \quad 0\leq \theta <\theta _0 \,, \quad \label{TD2a}\\ 
	\left. \frac{\partial u}{\partial \xi}\right| _{\xi =1+}=0\,,  \quad \theta _0<\theta \leq \pi  \,, \quad 
	\label{TD2b} \\
	\left. u\right| _{\xi \rightarrow \infty} \to 0\, \quad \mbox{for} \quad  0\leq
	\theta \leq \pi\, .   \quad \label{TD4}
\end{eqnarray}
Note that for the case under consideration when $\epsilon \to 0$ the condition at the outer boundary (\ref{LE2}) transforms into the so-called {\it regularity condition at infinity}  (\ref{TD4}). 

For the RCF (\ref{dm1na}) one, evidently, obtains
\begin{align}
	J\left( \theta_0; \epsilon\right) \to f \left( \theta_0 \right) \quad \mbox{as} \quad \epsilon \to 0 \,, \label{asthet1} 
\end{align}
where is the so-called {\it effective steric factor} (ESF) first introduced by Solc and Stockmayer in 1971.~\cite{Stockmayer71} Since that the ESF  has long been used in theoretical models of the reaction-diffusion processes with chemical anisotropy.
Particularly for the one-patch Solc-Stockmayer model the ESF takes the limiting values
\begin{align}
	f\left( \theta_0 \right) \to \left\{
	\begin{array}{ll}
		0 & \quad \mbox{as} \quad \theta _0 \to 0+ \, , \\
		1 & \quad \mbox{as} \quad \theta _0 \to \pi- \,
	\end{array}
	\right. \label{Bi3c} 
\end{align}
which correspond to the {\it fully reflecting} and {\it fully absorbing} reactive particles, respectively.

Note in passing that in this case at issue some authors considered a simple {\it heuristic geometric ESF} $f_g\left( \theta_0 \right)$
defined by~\cite{Piazza07,Lukzen08,Lukzen16}
\begin{eqnarray}
	f_g\left( \theta_0 \right)= \sqrt{{S_a}\left(\theta_0 \right)/{S}} \qquad \qquad \qquad \nonumber \\	
	\qquad= \frac{1}{\sqrt{2}}\sqrt{1-\cos \theta_0 } \sim \frac{1}{2}\theta_0\quad \mbox{as} \quad \theta_0 \to 0 \,, \label{Luk1}  
\end{eqnarray}
where $S_a\left(\theta_0 \right)/S$ represents the absorbing surface fraction of the
anisotropic sphere.~\footnote{Note in passing that in Ref.~\onlinecite{Lukzen08} reference~12 should be replaced by Ref.~\onlinecite{Traytak95} of the present work.} 
However, it turned out that this formula gives $57\%$ error compare to the exact 
small angle asymptotics~\cite{Traytak95}
\begin{align}
	f \left( \theta_0 \right)  \sim \frac{1}{\pi} \theta_0\quad \mbox{as} \quad \theta_0 \to 0  \label{Luk2}
\end{align}
derived for the first time in Ref.~\onlinecite{Lukzen81}. 
It is remarkable that Eq. (\ref{Luk2}) coincides with the well-known Hill formula (\ref{Hill75}).

The external mixed Dirichlet-Neumann problem (\ref{TD1})-(\ref{TD4}) was thoroughly studied previously, so we refer the interested reader to Ref.~\onlinecite{Traytak95} for details. Moreover,  all results for this case straightforwardly follow from the general theory developed below as $\epsilon \to 0$.

\section{General solution of the problem}\label{sec:GenSol}

Taking into account the linearity of the diffusion problem (\ref{LE1})-(\ref{MBC2}), it is expedient to use the {\it general linear superposition principle} and
look for the general solution of (\ref{LE1})-(\ref{MBC2}) as a decomposition of the {\it partial solutions} given in relevant partial domains.~\cite{Poster13,Grebenkov19} 
Really let $u^+_0: \Omega_{\xi}^{0} \to \left( 0,1\right)$, $ u^-: \Omega_{\xi}^{-} \to \left( 0,1\right)$
be harmonic functions in domains $\Omega_{\xi}^{0} $ and $\Omega_{\xi}^{-}$. 
Then in their intersection  $\Omega_{\xi}^{+}=\Omega_{\xi}^{0} \cap \Omega_{\xi}^{-}$ (\ref{Resc4}) we, evidently, have
\begin{eqnarray}
	u: \Omega_{\xi}^{+} \to \left( 0,1\right)\,,  \nonumber \\	
	u\left(\xi, \theta \right)=u^+_0\left(\xi, \theta \right)+u^-\left(\xi, \theta \right) \,.\label{GenS}
\end{eqnarray}
By means of Green’s representation of harmonic functions it may be directly shown that decomposition (\ref{GenS}) is unique.~\cite{Poster13,Galanti16a,Traytak18,Grebenkov19,Piazza19}

\begin{figure}[h!]
	\centering
	\includegraphics[scale=0.40]{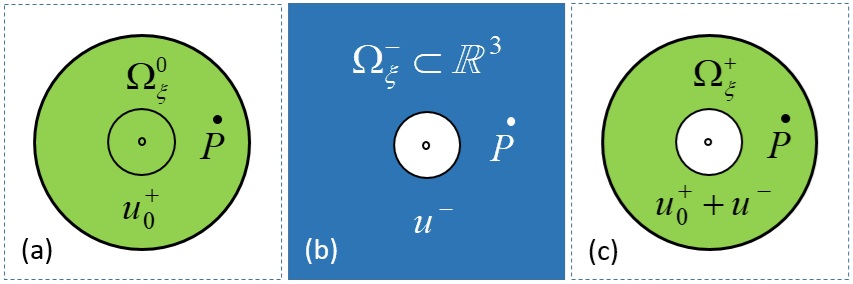}
	\caption{Schematic description of the decomposition procedure (\ref{GenS}): (a) solution $u^+_0$ inside the cavity sphere $\Omega_{\xi}^{0}$ (green); (b) solution $u^-$ outside the RP $\Omega_{\xi}^{-}$ (blue); (c) general solution  $u=u^+_0+u^-$  in the intersection $\Omega_{\xi}^{+}=\Omega_{\xi}^{0} \cap \Omega_{\xi}^{-}$ (green).}
	\label{v3}
\end{figure}

Thus, like for the case of two RPs~\cite{Traytak18,Piazza19} or two active sites on an inert sphere,~\cite{Barzykin07} located in $\Bbb{R}^3$ for the present case we deal with two partial solutions: $u^+_0$ due to cavity wall and $u^-$ caused because of the RP. 
We call  attention to the fact that the decomposition (\ref{GenS}) can be clearly illustrated by Fig.~\ref{v3}.

In its turn, any axially symmetric harmonic function regular inside the cavity sphere $\Omega_{\xi}^{0} 
$ and outside the RP $\Omega_{\xi}^{-}$ can be expanded on the functional {\it basis solutions} $\{\psi _l^{\pm }(\xi, \theta)\}_{l=0}^\infty $ given as follows: 
\begin{eqnarray}
	\psi _l^{+}\left(\xi, \theta \right):=\xi^lP_l\left( \cos \theta \right) \quad \mbox{in} \quad \Omega_{\xi}^{0} \,,\;\;  \label{Psi1}
	\\	
	\psi _l^{-}\left(\xi, \theta \right):=\xi^{-l-1}P_l\left( \cos \theta \right) \quad \mbox{in} \quad \Omega_{\xi}^{-}\,,  \label{Psi2}
\end{eqnarray}
where $P_l\left( \cos \theta \right) $ is the Legendre polynomial of degree $l$.
The subscripts "$+$" and "$-$" are commonly used to indicate {\it regular and irregular scalar axially symmetric solid  harmonics}, respectively.

Thus, the general axially symmetric solution to Eq. (\ref{Lap1}) owing to linearity of the problem reads~\cite{Traytak24}
\begin{eqnarray}
	u^+_0 \left(\xi, \theta \right) =\sum_{l=0}^{\infty} A_l^{+} \psi _l^{+}\left(\xi, \theta \right) \quad \mbox{in} \quad \Omega_{\xi}^{0} \,, \quad  \label{SolIn}\\
	u^-_1 \left(\xi, \theta \right) =\sum_{l=0}^{\infty} A_l^{-} \psi _l^{-}\left(\xi, \theta \right) \quad \mbox{in} \quad \Omega_{\xi}^{-} \,. \quad  \label{SolOut}
\end{eqnarray}
Here (\ref{SolIn}) and (\ref{SolOut}) are absolutely and uniformly convergent series expansions of regular and irregular
spherical harmonics. In turn, the associated coefficients $ \{A_l^{+}\}_{l=0}^\infty $ and $ \{A_l^{-}\}_{l=0}^\infty $ are called  {\it internal} and {\it external multipole} ($2^l${\it -pole}) {\it  moments}, respectively.

For both regular and irregular axially symmetric solid harmonics restriction to the unit sphere yields 
\begin{align}
	\left. \psi _l^{\pm}\left(\xi, \theta \right)\right| _{\xi =1+}= P_l\left(\cos\theta \right)\,. \label{PolL2}
\end{align}
Recall that Legendre's polynomials form an orthogonal basis $\{P_l\left( \cos \theta \right)\}_{l=0}^\infty $  in a Hilbert space $ L_2(0, \pi ) $ on the unit 3D sphere with the orthonormality condition and norm, respectively:
\begin{eqnarray}
	\int_{0}^{\pi} \sin\theta d\theta P_m\left(\cos\theta \right)P_l\left(\cos\theta \right) = \Vert P_l\Vert_{L_2}^2\delta_{lm}\,, \label{PolL3}	\\	
	\Vert P_l\Vert_{L_2}=\sqrt{\frac{2}{2l+1}}\,, \quad  l=\overline{0,\infty}\,, \nonumber
\end{eqnarray}
where $\delta_{lm}$ is the Kronecker delta which is given by $\delta_{lm} =0$ if $l \ne m$ and  $\delta_{lm} =1$ if $l = m$.

By taking into account expansion (\ref{GenS}) a general expression for the local concentration can be written as
\begin{align}	
	u\left(\xi, \theta \right) = \sum_{l=0}^\infty \left( A_l^{+} \xi^l + \frac{A^{-}_l}{\xi^{l+1}}\right) P_l\left(\cos \theta \right)	\,. \label{GenSol1}
\end{align}

It is clear that the use of this expression in the integral (\ref{dm1n}) gives
\begin{align}
	J\left( \theta_0, \epsilon \right) = A^{-}_0\,.  \label{Bi3a}
\end{align}
Thus, our main objective is to derive this RCF, i.e. monopole moment $  A^{-}_0 $.

Finally note that regardless of the parameter $\epsilon$ one has
\begin{align}
	J\left( \theta_0; \epsilon\right) \to 0 \quad \mbox{as} \quad \theta_0 \to 0\, . \label{asthet2}
\end{align}
The limit (\ref{asthet2}) is evident because for $\theta_0 \to 0$ local flux (\ref{MBC2}) and, therefore, integral (\ref{Bi3a}) becomes to be zero.

\section{Reduction to the dual series relations}\label{sec:RedDSR}

It is well known that the exact analytical solution to the proper mixed boundary value problem is often difficult if not impossible to obtain because of their complexity.~\cite{Sneddon66,Price07,Duffy08}
Moreover, they are seldom amenable to analytical treatment including asymptotic analysis and that is why analytical or at least semi-analytical solution of any mixed boundary value problem is of great interest.

Nevertheless for the problem under consideration it is possible to derive the so-called {\it dual series relations} (DSR)~\cite{Sneddon66} or {\it dual Fourier-Legendre series}~\cite{Duffy08} using separation variables approach. It is noteworthy that in turn the DSR (\ref{DSR1a}), (\ref{DSR2a}) can be solved, at least semi-analytically, either by reducing to an appropriate infinite set of equation or Fredholm integral equation of the second kind.~\cite{Minkov64,Sneddon66,Tachiya95a,Duffy08}

Substitution of Eq. (\ref{GenSol1}) into condition on the outer boundary (\ref{LE2}) and mixed boundary conditions (\ref{MBC1}), (\ref{MBC2}) yields 
\begin{eqnarray}
	\sum\limits_{l=0}^\infty \left( A_l^{+} \epsilon^{-l} + {A^{-}_l}{\epsilon^{l+1}}\right) P_l\left( \cos \theta \right) = 0 \,, \quad 0\leq \theta <\pi \,, \qquad \label{LE9v}\\
	\sum\limits_{l=0}^\infty \left( A_l^{+}  + {A^{-}_l}\right) P_l\left( \cos \theta \right) = 1 \,, \quad 0\leq \theta <\theta _0\,, \qquad \label{LE9}\\ 
	\sum\limits_{l=0}^\infty \left[ lA_l^{+} - (l+1)A_l^{-} \right] P_l\left( \cos \theta
	\right) =0\,, \quad \theta _0<\theta \leq \pi \,. \qquad \label{LE10}
\end{eqnarray}

It is a simple matter to eliminate here internal moments $A^{+}_l$ in order to reduce obtained three series relations   (\ref{LE9v})-(\ref{LE10}) to the corresponding system of two equations with respect to only one infinite sequence of unknown external moments $ \{A_l^{-}\}_{l=0}^\infty $. 

It follows directly from Eq. (\ref{LE9v}) that the connection holds
\begin{align}
	{A^{+}_l} = -  {A^{-}_l} \epsilon^{2l+1}   \quad \mbox{for all} \quad  l=\overline{0,\infty}\,. \label{LE9t}
\end{align}
Note in passing that for $ \epsilon \to 0 $ we, obviously, have $ {A^{+}_l} \to 0 $ for all  $l=\overline{0,\infty}$ as it should be.

Thus, the problem (\ref{LE9v})-(\ref{LE10}) is readily transformed to the following DSR 
\begin{eqnarray}
	\sum\limits_{l=0}^\infty \left( 1  - \epsilon^{2l+1}  \right)  A_l^{-} P_l\left( \cos \theta \right) =  1 \,, \quad 0\leq \theta <\theta _0 \,, \qquad \label{DSR1a}\\
	\sum\limits_{l=0}^\infty \left(1+ l + l\epsilon^{2l+1} \right) A_l^{-}  P_l\left( \cos \theta
	\right) = 0\,, \quad \theta _0<\theta \leq \pi \,. \qquad  \label{DSR2a}  
\end{eqnarray}
Moreover, obtained connection (\ref{LE9t}) allows us to simplify the expression (\ref{GenSol1}) for the local concentration
\begin{align}	
	u\left(\xi, \theta \right) = \sum_{l=0}^\infty \left( \frac{1}{\xi^{l+1}}-  \epsilon^{2l+1} \xi^l\right)A^{-}_l P_l\left(\cos \theta \right)\,. \label{GenSol1a}
\end{align}

Then to solve Eqs. (\ref{DSR1a}), (\ref{DSR2a}) one should reduce them to their so-called {\it canonical DSR}.~\cite{Sneddon66,Traytak95} For this purpose we introduce new unknown sequence of coefficients $ \{X_l\}_{l=0}^\infty $ by the relation
\begin{align}
	\left(1+ l + l \epsilon^{2l+1} \right) A_l^{-} = \left(l +\frac{1}{2} \right)X_l\,.	\label{Rel1}
\end{align}
After some simple algebra the DSR (\ref{DSR1a}), (\ref{DSR2a}) may be transformed to 
\begin{align}
	\sum\limits_{l=0}^\infty\left[1 - q_l(\epsilon) \right] X_lP_l\left(\cos \theta \right) =  1 \,, \quad 0\leq \theta <\theta _0 \,, \label{CanDSR1}\\
	\sum\limits_{l=0}^\infty \left(l +\frac{1}{2} \right)X_l P_l\left(\cos \theta \right) = 0 \,, \quad \theta _0<\theta \leq \pi \,, \label{CanDSR2}  
\end{align}
where 
\begin{eqnarray}
	q_l(\epsilon):= w_l(\epsilon) +  \left[1 - w_l(\epsilon) \right]\epsilon^{2l+1} \,, \qquad \label{LE10t}\\
	w_l(\epsilon) =   \frac{ 1 + 2 l \epsilon^{2l+1}}{2\left( 1+ l + l \epsilon^{2l+1} \right)} \,. \qquad  \label{LE11t}  
\end{eqnarray}

It is evident that 
\begin{eqnarray}
	q_0(\epsilon):=  \frac{1}{2}\left( 1 + \epsilon \right) \,, \quad \label{Lam0}\\
	w_l(\epsilon)  \to  1/{2(l+1)}=q_l(0)\,, \quad l \ge 0 \quad \mbox{as} \quad \epsilon \to 0 \,. \quad \label{Inf1}  
\end{eqnarray}

One can see that 
\begin{align}  
	A_l^{-} = \left[1 - w_l(\epsilon) \right] X_l \label{AXcon}
\end{align}
and using relation (\ref{Bi3a}) one arrives at the important formula for the correction factor
\begin{align}
	J\left( \theta_0, \epsilon \right) = \frac{1}{2} X_0\left( \theta_0, \epsilon \right) \,. \label{ESF0}
\end{align}
Therefore, to calculate the required rate (\ref{dm1na}) we only need to know the zeroth element of the sequence  $ \{X_l\} $ (in what follows we will use this very notation for short).

In its turn utilizing connection (\ref{AXcon}) expression for the local concentration (\ref{GenSol1a}) it is expedient to represent by means of solution $ \{X_l\} $ to the canonical DSR (\ref{CanDSR1}), (\ref{CanDSR2}) as follows:
\begin{eqnarray}
	u\left(\xi, \theta \right) = \sum_{l=0}^\infty \left( \frac{1}{\xi^{l+1}}-  \epsilon^{2l+1} \xi^l\right) \nonumber \\
	\times 	\left[1 - w_l(\epsilon) \right] X_l P_l\left(\cos \theta \right)\,. \label{GenSol1b} 
\end{eqnarray}

\section{Reduction to an infinite system of linear equations}\label{sec:RedISLAE}

Following Minkov~\cite{Minkov64} we will reduce the canonical DSR (\ref{CanDSR1}),  (\ref{CanDSR2}) to so-called {\it resolving infinite set of linear algebraic equations} (ISLAE).~\footnote{Since in this paper only resolving infinite systems are studied, below we omit the qualifier {\it resolving} for brevity.}  For this goal first let us consider an auxiliary DSR of the form
\begin{eqnarray}
	\sum\limits_{l=0}^\infty Y_lP_l\left(\cos \theta \right) =  g\left(\theta \right)  \,, \quad 0\leq \theta <\theta _0 \,, \label{ExS1}\\
	\sum\limits_{l=0}^\infty \left(l +\frac{1}{2} \right)Y_l P_l\left(\cos \theta \right) = 0 \,, \quad \theta _0<\theta \leq \pi \,, \label{ExS2}  
\end{eqnarray}
where $g\left(\theta \right)$ is known arbitrary continuous function given in the interval $\left(0,\theta_0 \right)$ and real number sequence $ \{Y_l\} $ is to be determined.
It is well known that these DSR possesses the exact solution~\cite{Sneddon66} 
\begin{align}
	Y_l=\frac{\sqrt{2}}\pi \int\limits_0^{\theta _0}dt\cos \left[\left(l +\frac{1}{2} \right)t \right]\frac d{dt}\int\limits_0^td\tau \frac{g\left( \tau \right)
		\sin \tau }{\sqrt{\cos \tau -\cos t}}.  \label{Bi25}
\end{align}
Clearly Eq. (\ref{CanDSR1}) may be represented in the self-consistent form 
\begin{align}
	\sum\limits_{l=0}^\infty X_lP_l\left(\cos \theta \right) =G\left( \theta
	\right), \quad \theta _0<\theta \leq \pi,  \label{Bi26}
\end{align}
where 
$$
G\left( \theta \right) =1
+\sum\limits_{l=0}^\infty q_lX_lP_l\left(\cos \theta \right) \,.
$$
Utilizing exact formula (\ref{Bi25}) we find the implicit representation of the solution to the canonical DSR (\ref{CanDSR1}),  (\ref{CanDSR2})
\begin{align}
	X_l=\frac{\sqrt{2}}\pi \int\limits_0^{\theta _0}dt\cos \left[\left(l +\frac{1}{2} \right)t \right]\frac d{dt}\int\limits_0^td\tau \frac{G\left( \tau \right)
		\sin \tau }{\sqrt{\cos \tau -\cos t}}\,.  \label{Bi27}
\end{align}
The integrals appearing here may be computed explicitly.
Indeed one can calculate the inner integral in Eq. (\ref{Bi27}) by use of the relation for the Legendre polynomials of degree $m$  ($m \ge 0$)~\cite{Ryzhik15}
\begin{align}
	\int_0^u\frac{P_m(\cos \theta )\sin \theta \,d\theta }{\sqrt{\cos \theta
			-\cos u}}=\sqrt{2}\Vert P_m\Vert_{L_2}^2 \sin \left[ \left( m+\frac 12\right) u\right]\,.  \label{GradRyz}
\end{align}
Then the right hand side of Eq. (\ref{Bi27}) becomes an integral that involve trigonometric functions.

Finally we get the desired ISLAE: 
\begin{align}
	X_l-\sum\limits_{m=0}^\infty M_{lm}X_m=B_l\,,\quad l=\overline{0,\infty} 	\,.  \label{Bi28}
\end{align}
Hereafter infinite matrix is
\begin{align}
	M_{lm}\left( \theta _0; \epsilon\right):=q_m(\epsilon)Q_{lm}\left( \theta _0\right)\,,
	\label{Bi28s}
\end{align} 
where matrix function reads
\begin{eqnarray}
	Q_{lm}\left( \theta\right)= \frac 1\pi \left\{ \frac{\sin \left[ \left(
		l+m+1\right) \theta\right] }{\left( l+m+1\right) }  \right.  \qquad \qquad \nonumber \\
	\left. +\frac{\sin \left[
		\left( l-m\right) \theta \right] }{\left( l-m\right) }\left( 1-\delta
	_{lm}\right) +\theta \delta _{lm} \right\} 	\label{Bi29}
\end{eqnarray}
and $B_l=X_l^{0}\left( \theta _0\right):=Q_{l0}\left( \theta _0\right)$.

Solution to the ISLAE (\ref{Bi28}) is the sequence $ \{X_l\} $ and when $ \theta _0 =\pi $ we have $ Q_{lm}\left( \pi\right) = \delta _{lm} $ and, therefore, this system degenerates to $ X_0 = 2 $ and $ X_l = 0 $ if $ l \in \Bbb{N}$. Thus at the ends of the interval $\left(0, \pi\right)$ ISLAE (\ref{Bi28}) and Eq. (\ref{ESF0}) yield
\begin{align}
	J\left( \theta_0; \epsilon \right) \to \left\{
	\begin{array}{ll}
		0 & \quad \mbox{as} \quad \theta _0 \to 0+ \, , \\
		\left(1-\epsilon\right)^{-1} & \quad \mbox{as} \quad \theta _0 \to \pi- \,,
	\end{array}
	\right. \label{Lim1} 
\end{align}
which is to be expected.

\begin{remark}
	It is significant that there is another method to solve the canonical form of the DSR (\ref{CanDSR1}), (\ref{CanDSR2})  by reduction of these relations to a Fredholm integral equation of the second kind.~\cite{Minkov60,Tachiya95a}
\end{remark}
Note also that in Ref.~\onlinecite{Kovalenko86} Aleksandrov and Kovalenko proposed quite different method to reduce some problems on the continuum mechanics with mixed boundary conditions, particularly for Laplace's equation, to the appropriate ISLAE.

\section{Solution of the ISLAE}\label{sec:SolISLAE}

Various problems of mathematical physics and its applications lead to ISLAE which are therefore of paramount importance.

The most basic method for solving an ISLAE is the so-called {\it reduction method}, in which the infinite system is truncated and then a finite-size system is solved by standard methods of linear algebra.  However, the solution of the truncated system may or may not converge to that of the ISLAE so that the questions of existence and uniqueness of the solution are particularly important.  

Study of the ISLAE in appropriate classical sequence spaces fills a highly important place in the DSR method and, contrary to the the case of finite system of linear algebraic equations, the primary questions of existence and uniqueness of the solution of the ISLAE is a typical problem of functional analysis. So, in the next two Subsections, following well-known books Refs.~\onlinecite{Kantorovich64,Kantorovich82,Kovalenko86}, we present a few important facts concerning solution to the ISLAE.

\subsection{Basic definitions} \label{subsec:BasDef}

We start with a general important
\begin{definition}
	For given infinite $\left(\overline{0,\infty}\right) \times \left(\overline{0,\infty}\right)$ matrix with elements $M_{lm}\in \Bbb{R}$ and inhomogeneous term $\{B_l\}$ the system (\ref{Bi28})
	is called the ISLAE (with respect to the unknown elements $X_l \in \Bbb{R}$, $l=\overline{0,\infty}$) of the second kind.
\end{definition}

Let us recall now the definition of a solution to the ISLAE (\ref{Bi28})
\begin{definition}
	If for an infinite matrix $(M_{lm})$ defined by (\ref{Bi28}), there exists an infinite sequence $\{X_m\}$ such that the series
	\begin{equation*}
		\sum\limits_{m=0}^\infty (\delta_{lm}-M_{lm}) X_m  
	\end{equation*} 
	converge to $B_l$ for all $l=\overline{0,\infty}$, then the sequence $\{X_m\}$ is termed a solution of the ISLAE (\ref{Bi28}).
\end{definition}

\subsection{Solvability of the ISLAE by truncation method} \label{subsec:Solvabil}

\begin{definition} \label{Tranc1}
	A finite system of $n$ linear algebraic equations, based on $B_l$ and
	$M_{lm}$ with $l,m = \overline{1,n}$,
	\begin{equation}
		X_l^{(n)} - \sum\limits_{m=0}^n M_{lm} \, X_m^{(n)} = B_l\,,  \quad l\in \overline{0,n}  \label{Bi31}
	\end{equation}
	is called a truncation of the ISLAE (\ref{Bi28}) to the order $n$.~\cite{Kantorovich64}
\end{definition}
The number $n$ can be also called the {\it order of truncation approximation}.

Let us proceed to the description of the criteria of solvability of
the ISLAE.

\begin{definition} \label{LoInt1}
	The ISLAE is termed regular if 
	\begin{align}
		\sum\limits_{m=0}^\infty \vert M_{lm}\vert <1 \,,\quad l=\overline{0,\infty}\,. \label{Qlm6a}  
	\end{align}
\end{definition}
However the condition (\ref{Qlm6a}) is rather strong and can be weakened to solve ISLAE by the reduction method.

\begin{lemma} \label{Lemma2}
	The regularity condition (\ref{Qlm6a}) is sufficient for the ISLAE (\ref{Bi28}) to be solved by truncation (\ref{Bi31}) such that
	\begin{align}
		\lim_{n \rightarrow \infty }X_l^{( n)}=X_l \quad \mbox{for all} \quad l=\overline{0,\infty} \label{Lemma2}  
	\end{align}
	is the solution of the ISLAE (\ref{Bi28}).
\end{lemma}
The proof of Lemma~\ref{Lemma2} can be found e.g., in Ref.~\onlinecite{Kantorovich64}.

\begin{definition}
	Let $ \ell_\infty$ defined to be the space of all bounded sequences $\{B_l\}$, i.e., $\vert B_l \vert \le C_b >0$ for all $l=\overline{0,\infty}$. Space $ \ell_\infty$ is a  Banach space  endowed with the norm: $\Vert B_l\Vert_\infty:=\sup_{l \ge 0 }\vert B_l \vert$
\end{definition}

It is clear from Definition~\ref{LoInt1} that the inequality holds
\begin{align}
	\sum\limits_{m=0}^\infty \vert M_{lm}\vert \leq \Vert M \Vert_{\infty} \,,\quad l=\overline{0,\infty}\,, 
	\label{Qlm6b}  
\end{align}
where the matrix norm is defined as
\begin{align}
	\Vert M \Vert_{\infty}\left( \theta_0 ; \epsilon \right):= \sup_{l} \sum\limits_{m=0}^\infty \vert M_{lm}\left( \theta_0 ; \epsilon \right)\vert \,.\label{NormIn}  
\end{align}
So the following condition 
\begin{align}
	\Vert M \Vert_{\infty}\left( \theta_0 ; \epsilon \right) < 1 \label{Qlm7}  
\end{align}
is sufficient for the system (\ref{Bi28}) to be solved by truncation method. 

\begin{thm} \label{Theorem1}
	(Existence and Uniqueness) If the inhomogeneous term $\{ B_l\} \in \ell_\infty$, the regular ISLAE (\ref{Bi28}) has a unique
	solution  $\{X_m\} \in \ell_\infty$.
\end{thm}
The proof of Theorem~\ref{Theorem1} is based on the contraction mapping principle for corresponding matrix operators.

Note that the regularity condition (\ref{Qlm6a}) is sufficient to solve
ISLAE by both simple reduction method and iterations.~\cite{Kantorovich82}

The following assertion holds true
\begin{lemma} \label{Lemma1}
	The inhomogeneous term	$X_l^{0}$ in the ISLAE (\ref{Bi28}) belongs to the space of bounded sequences, that is $\{X_l\} \in \ell_\infty$.
\end{lemma}
The proof of the Lemma~\ref{Lemma1} implies directly from the evident estimates
$$
0 \le X^0_0 \le 1\,, \quad \vert X^0_l \vert < 3/2 \pi \quad \mbox{for all} \quad l\in \Bbb{N}\,.
$$

\begin{remark}
	We emphasize that theory of the ISLAE in space of bounded sequences $ \ell_\infty$ may be readily extended to the theory when solvability condition by the reduction method appeared to be considerably weaker (see Ref.~\onlinecite{Traytak19} and the references therein)
	\begin{align}
		\sum\limits_{l,m=0}^\infty  M^2_{lm} <+\infty \,, \label{HSL2a} 
	\end{align}
	where $ \ell_2$ is the Hilbert space of sequences $\{X_m\}$ such that
	$$
	\{X_m\} \in \ell_2 :=\{X_m \in \Bbb{R}: \sum\limits_{m=0}^\infty  X^2_{m} <+\infty \}\,.
	$$
\end{remark}

\subsection{Approximate analytical solution}

Note that the above approximations the index $n$ corresponds to the number of equations in the truncated set of equations (\ref{Bi31}).  

The infinite system (\ref{Bi28}) may be truncated (see Definition~\ref{Tranc1}) and solved by iteration. Thus one obtains for the zeroth and first order approximations of the zero element $X_0$ of the sequence $\{X_l\}$
\begin{eqnarray}
	X_0 \approx X_0^{(i)} \,,\;\; (i=0,1) \,,   \qquad \nonumber\\ 
	X_0^{(0)}=X_0^{0}\,, \quad X_0^{(1)} =\frac{X_0^{0}}{\left(1-  M_{00}\right)}\,. \label{Appr1} 
\end{eqnarray} 

Thus, with the help of Eq. (\ref{ESF0}) one obtains for the zeroth order approximation of RCF
\begin{eqnarray}
	J\left( \theta_0; \epsilon \right) \approx J^{(0)}\left( \theta_0 \right) =f^{(0)}\left( \theta_0 \right)\,, \label{Zero} \\ 
	f^{(0)}\left( \theta_0 \right):= \frac {1}{2\pi}\left(\theta_0+\sin \theta_0 \right)\,.\; \label{Appr1} 
\end{eqnarray} 
Notably, the zeroth order approximation of RCF $J^{(0)}\left( \theta_0 \right)$ is independent of the thickness ratio $ \epsilon $ (\ref{thickness2}). 
At the same time formula (\ref{Zero}) is a small angle $\theta_0$ approximation, which coincides with that as $\epsilon \to 0$.~\cite{Traytak95}

Similarly one readily finds the first order approximation. Note that dependence of the RCF on parameter $ \epsilon $ starts from the first order approximation.

The use of Eq. (\ref{Lam0}) gives
$$
M_{00}\left( \theta _0; \epsilon\right)= q_0(\epsilon)Q_{00}\left( \theta _0\right)= \left( 1 + \epsilon \right)f^{(0)} \left(\theta_0 \right) \,.
$$

Comparison of results calculated for the correction factor

The first order approximation for the RCF follows from Eqs. (\ref{ESF0}) and (\ref{Appr1}) 
\begin{align}
	J\left( \theta_0; \epsilon \right) \approx J^{(1)}\left( \theta_0; \epsilon \right) =\frac{f^{(0)}\left( \theta_0 \right)}{1-  \left( 1 + \epsilon \right)f^{(0)}\left( \theta_0 \right)}\,, \label{App2}  
\end{align}
where $J^{(1)}\left( \theta_0; \epsilon \right)$ is the first approximation of the RCF.
The first approximation of the RCF (\ref{App2}) can be readily transformed to a more convenient form
\begin{eqnarray}
	J^{(1)}\left( \theta_0; \epsilon \right) =\frac{f^{(1)}\left( \theta_0 \right)}{1- \epsilon f^{(1)}\left( \theta_0 \right)}\,, \label{App2a} \\ 
	f^{(1)}\left( \theta_0 \right) =  \frac{f^{(0)}\left( \theta_0 \right)}{1-f^{(0)}\left( \theta_0 \right)}\,.\; \label{Appr2b} 
\end{eqnarray}

Clearly function $J^{(1)}\left( \theta_0; \epsilon \right)$ is increasing with respect to the thickness ratio $\epsilon \in (0, 1)$. Such kind of behavior is entirely consistent with the above considered limiting case of ideally absorbing chemically isotropic  RP (\ref{SSE4}) and by the same reason physically this fact stems due to diffusive interaction effects.~\cite{Piazza19,Traytak24} 
For $\epsilon \to 0$ the diffusive interaction disappears and, therefore, we can reproduce the result of our previous work~\cite{Traytak95}
\begin{align}
	J^{(1)}\left( \theta_0; \epsilon \right) \to f^{(1)}\left( \theta_0 \right)  \,.\label{App3}  
\end{align}

It turns out that the first approximation of the RCF $J^{(1)}$ (\ref{App2a}) possesses a few amazing properties.
\begin{itemize}
	\item  For all values  $\left( \theta_0, \epsilon\right) \in \left( 0, \pi \right)\times \left(0, 1\right)$ the first approximation $J^{(1)}$ is a uniform lower bound of the exact RCF $J$, i.e.
	\begin{align}
		J^{(1)}\left( \theta_0; \epsilon \right)<J\left( \theta_0; \epsilon \right) \,. \label{LBoun}  
	\end{align}
	\item  $J^{(1)}$ reproduces also correct limits  (\ref{Lim1}) 
	\begin{align}
		J^{(1)}\left( \theta_0; \epsilon \right) \to \left\{
		\begin{array}{ll}
			0 & \quad \mbox{as} \quad \theta _0 \to 0+ \, , \\
			\left(1-\epsilon\right)^{-1} & \quad \mbox{as} \quad \theta _0 \to \pi- \,.
		\end{array}
		\right. \label{ApprL}
	\end{align} 
	\item  The small angle expansion of $J^{(1)}$, irrespective of the thickness ratio $\epsilon$ yields~\cite{Traytak95}
	\begin{align}
		J^{(1)}\left( \theta_0; \epsilon \right) \sim \theta_0/\pi \quad \mbox{as} \quad \theta _0 \to 0+ \,.
		\label{Hill}
	\end{align}
\end{itemize}
It seems surprising, but small angle asymptotics (\ref{Hill}) leads to the correct value even as $\theta _0 \to \pi-$.

Besides simple analytical approximations for the RCF derived in this section the  ISLAE may be easily solved numerically to yield both the RCF (\ref{ESF0}) and the local concentration profile (\ref{GenSol1a}) with a high accuracy.

\section{Numerical results}~\label{sec:NumRes}

To solve ISLAE (\ref{Bi28}) numerically we used the truncation method (see its description in subsection~\ref{subsec:Solvabil}), which in turn was implemented with codes from the NumPy library for the Python programming language.~\cite{Langtangen08} The solution converges rather fast with increasing truncation order $n$. For example, the maximum difference between orders 100 and 200 is on the order of $10^{-5}$.

We have proved in Sec.~\ref{sec:RedDSR} that to calculate the trapping rate (\ref{dm1na}) (or, equivalently, correction factor (\ref{ESF0})) it is only necessary to find the zeroth element of the solution to the ISLAE (\ref{Bi28}) $ X_0 $. So, first let us present our numerical results obtained for the trapping rate.

\subsection{Estimation of the trapping rate}~\label{subsec:trapRate}

Since the desired solution $\{X_l\}$ of the ISLAE (\ref{Bi28}) can be obtained numerically to any necessary degree of accuracy and, moreover, the convergence of this procedure is known to be quite fast one can treat numerical  computations as exact results.

\subsubsection{Exact results}~\label{subsec:trapRate}

\begin{figure*}[h]	
	\centering
	~\\
	~\\
	\includegraphics[width=0.59\textwidth]{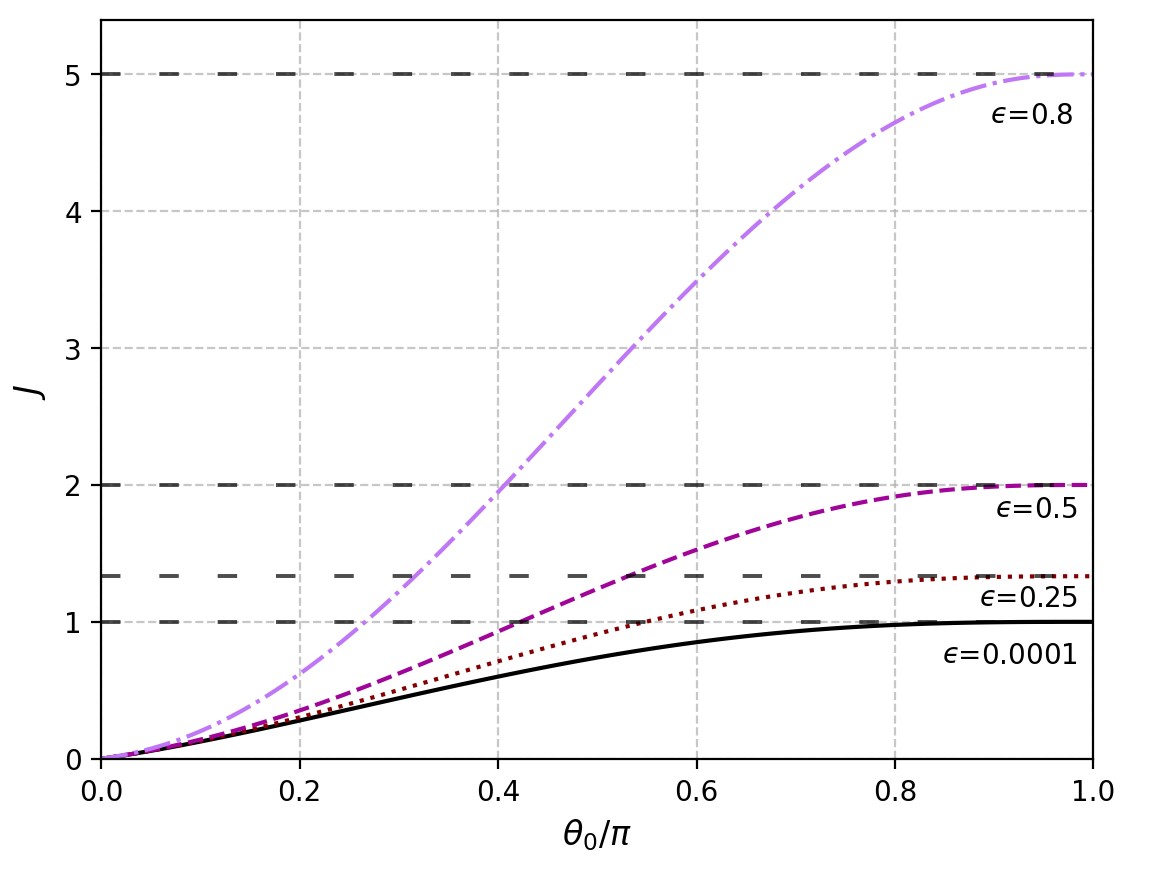}
	\caption{The rate correction factor $ J\left( \theta_0, \epsilon\right)$ as a function on the angular size $\theta_0$ at four different magnitudes of the thickness ratio: $\epsilon = 10^{-4};\, 0.25;\, 0.5;\, 0.8$. Straight dashed lines indicate appropriate horizontal asymptotes as $\theta_0 \to \pi-$ (\ref{Lim1}).}	
	\label{fig3}
\end{figure*}

Results calculated for the correction factor $J\left( \theta_0; \epsilon\right)$ at different thickness ratio $\epsilon$  are plotted in Fig.~\ref{fig3}. The significant growth of the diffusive interaction effects with increase of the thickness ratio $\epsilon$ and angular size of active site $\theta_0$ is clearly seen from the curves displayed in Fig.~\ref{fig3}. 

It is evident that horizontal asymptotes describe the spherically symmetric Berg's model $\theta_0 = \pi$ considered in subsection~\ref{subsec:Bergmodel}. As would be expected from Berg's model considerations for the case of small shell thickness $h$ (\ref{Bi28}) we have faced some difficulties during solution of the ISLAE.  
Indeed, the numerical calculations of the sup-norm (\ref{NormIn}) showed that in case of thin shells $h \ll 1$, for example, already at $h=0.2$ sufficient condition (\ref{Qlm7}) does not hold for the range:  $2.5 \lesssim \theta_0/\pi \lesssim 0.8$.

We reproduce numerical results obtained previously for unbounded domain~\cite{Traytak95} in Table~\ref{tab:table1}.

\begin{table*}[h!]
	\caption{\label{tab:table1} The correction factor $J\left( \theta_0; \epsilon\right)$ at different thickness ratio $\epsilon$.}
	\vspace{0.2cm}
	\begin{tabular}{|l|l|l|l|l|l|l|l|l|l|l|l|l|l|l|l|}
		\hline
		$\theta_0 / \pi$   & 0.067 & 0.133 & 0.200 & 0.267 & 0.333 & 0.400 & 0.467 & 0.533 & 0.600 & 0.667 & 0.733 & 0.800 & 0.867 & 0.933 & 1.000 \\ \hline
		$ \epsilon =   0$  & 0.080 & 0.174 & 0.279 & 0.389 & 0.496 & 0.600 & 0.696 & 0.780 & 0.851 & 0.908 & 0.950 & 0.978 & 0.993 & 0.999 & 1.000 \\ \hline
		$ \epsilon =   0.1$ & 0.081 & 0.178 & 0.288 & 0.404 & 0.522 & 0.639 & 0.748 & 0.846 & 0.931 & 0.999 & 1.049 & 1.083 & 1.103 & 1.110 & 1.111 \\ \hline
	\end{tabular}
\label{table1}
\end{table*}

It is significant that in Ref.~\onlinecite{McCammon05} to solve the steady-state diffusion problem for the unbounded domain case by means of a {\it finite element method} (see, e.g., Ref.~\onlinecite{Scott07}) the outer boundary condition (\ref{Seki5}) was posed at $R_0=40 R$. Thus, instead of real "infinity" (as $ \epsilon \to 0- $) the bulk concentration $c_B$ was prescribed at some {\it actual infinity}  $ \epsilon_{\infty}$, which was assumed to be $0.025$.

Meanwhile in Ref.~\onlinecite{Eun20}, using similar approach, to evaluate validity of numerical calculation 
Eun used the model of a spherical cavity treated the magnitude of the actual infinity  $ \epsilon_{\infty} =0.1 $. One can see from the second line of  Table~\ref{tab:table1} that at $R_0=10 R$ relative error attains its maximum at $\theta_0=\pi$ and becomes about $11.1\%$.
However, our numerical results given in Table~\ref{tab:table1} entirely confirm calculations performed by Eun for small-sized patches.

Note that to achieve the accuracy presented in Table~\ref{tab:table1} (as well as below in Table~\ref{tab:table2}), the truncation order cannot exceed 100. With this value of $n$, the computations were completed highly fast.

\subsubsection{Comparison with analytical approximations}~\label{subsec:trapRate}

\begin{figure}[h!]
	\centering{\includegraphics[width=0.52\textwidth]{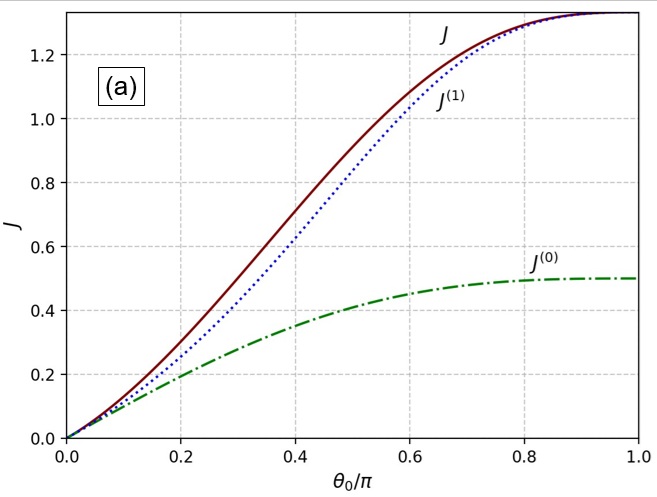}
		\includegraphics[width=0.52\textwidth]{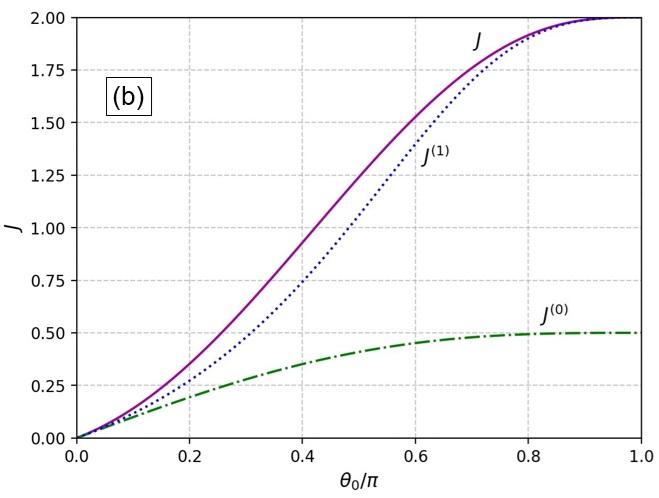}
		\includegraphics[width=0.52\textwidth]{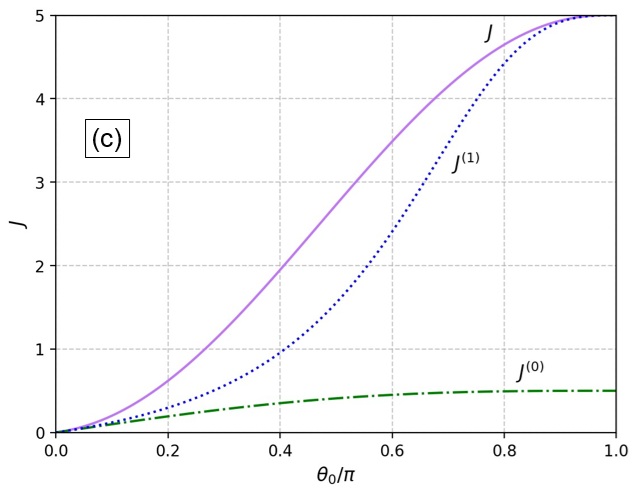}}
	\caption{The correction factor $J\left( \theta_0; \epsilon \right)$ (red solid curves) with its zeroth $J^{(0)}\left( \theta_0 \right)$ (green dash-dot curves) and first $J^{(1)}\left( \theta_0; \epsilon \right)$ (blue dotted curves) approximations as function of $\theta_0$ at: (a) $\epsilon = 0.25$, (b) $\epsilon = 0.5$ and (c) $\epsilon = 0.8$.}
	\label{fig8}
\end{figure}


Comparison between exact correction factor $J\left( \theta_0; \epsilon \right)$ (\ref{ESF0}) with its zeroth and first order approximations at various values of the thickness ratio $\epsilon$ is made in Fig.~\ref{fig8}.
Simple inspection the curves displayed in Fig.~\ref{fig8} clearly discloses that with the growth of the diffusive interaction (with an increase of the thickness ratio $\epsilon$) both correction factor approximations zeroth $J^{(0)}\left( \theta_0 \right)$ (\ref{Appr1}) and first $J^{(1)}\left( \theta_0; \epsilon \right)$ (\ref{App2a}) become worse. 

Since the zeroth approximation $J^{(0)}\left( \theta_0 \right)$  does not  describe diffusive interaction effects and does not possesses necessary property (\ref{Lim1}), furthermore we will consider the first approximation $J^{(1)}\left( \theta_0; \epsilon \right)$ only.
The role of the diffusive interaction effect is especially well highlighted from numerical results are tabulated in Table~\ref{tab:table2}).  

\begin{table*}[h!]
\caption{\label{tab:table2} The correction factor $J\left( \theta_0; \epsilon\right)$ and its first approximation $J^{(1)}\left( \theta_0; \epsilon\right)$ at different thickness ratio $\epsilon$.}
\vspace{0.2cm}
\begin{tabular}{|l|l|l|l|l|l|l|l|l|l|l|l|l|l|l|l|}
	\hline
	$\qquad \theta_0 / \pi$                     & 0.067 & 0.133 & 0.200 & 0.267 & 0.333 & 0.400 & 0.467 & 0.533 & 0.600 & 0.667 & 0.733 & 0.800 & 0.867 & 0.933 & 1.000 \\ \hline
	$ \quad J: (\epsilon = 0.25)$       & 0.082 & 0.183 & 0.303 & 0.434 & 0.571 & 0.712 & 0.848 & 0.973 & 1.084 & 1.177 & 1.246 & 1.294 & 1.321 & 1.332 & 1.333 \\ \hline
	$J^{(1)}: (\epsilon = 0.25)$ & 0.073 & 0.157 & 0.255 & 0.368 & 0.491 & 0.627 & 0.768 & 0.906 & 1.036 & 1.147 & 1.231 & 1.288 & 1.320 & 1.332 & 1.333 \\ \hline
	$\quad J: (\epsilon = 0.5)$        & 0.086 & 0.203 & 0.353 & 0.529 & 0.722 & 0.930 & 1.139 & 1.339 & 1.527 & 1.690 & 1.820 & 1.916 & 1.973 & 1.996 & 2.000 \\ \hline
	$J^{(1)}: (\epsilon = 0.5)$  & 0.074 & 0.163 & 0.273 & 0.405 & 0.560 & 0.743 & 0.950 & 1.171 & 1.398 & 1.608 & 1.778 & 1.901 & 1.970 & 1.996 & 2.000 \\ \hline
	$\quad J: (\epsilon = 0.8)$        & 0.111 & 0.317 & 0.620 & 1.006 & 1.451 & 1.949 & 2.470 & 2.985 & 3.487 & 3.945 & 4.334 & 4.648 & 4.866 & 4.978 & 5.000 \\ \hline
	$J^{(1)}: (\epsilon = 0.8)$  & 0.076 & 0.172 & 0.297 & 0.461 & 0.673 & 0.956 & 1.329 & 1.805 & 2.407 & 3.105 & 3.812 & 4.422 & 4.815 & 4.975 & 5.000 \\ \hline
\end{tabular}
\label{table2}
\end{table*}

\begin{figure*} [h!]
	\centering
	\includegraphics[width=0.60\textwidth]{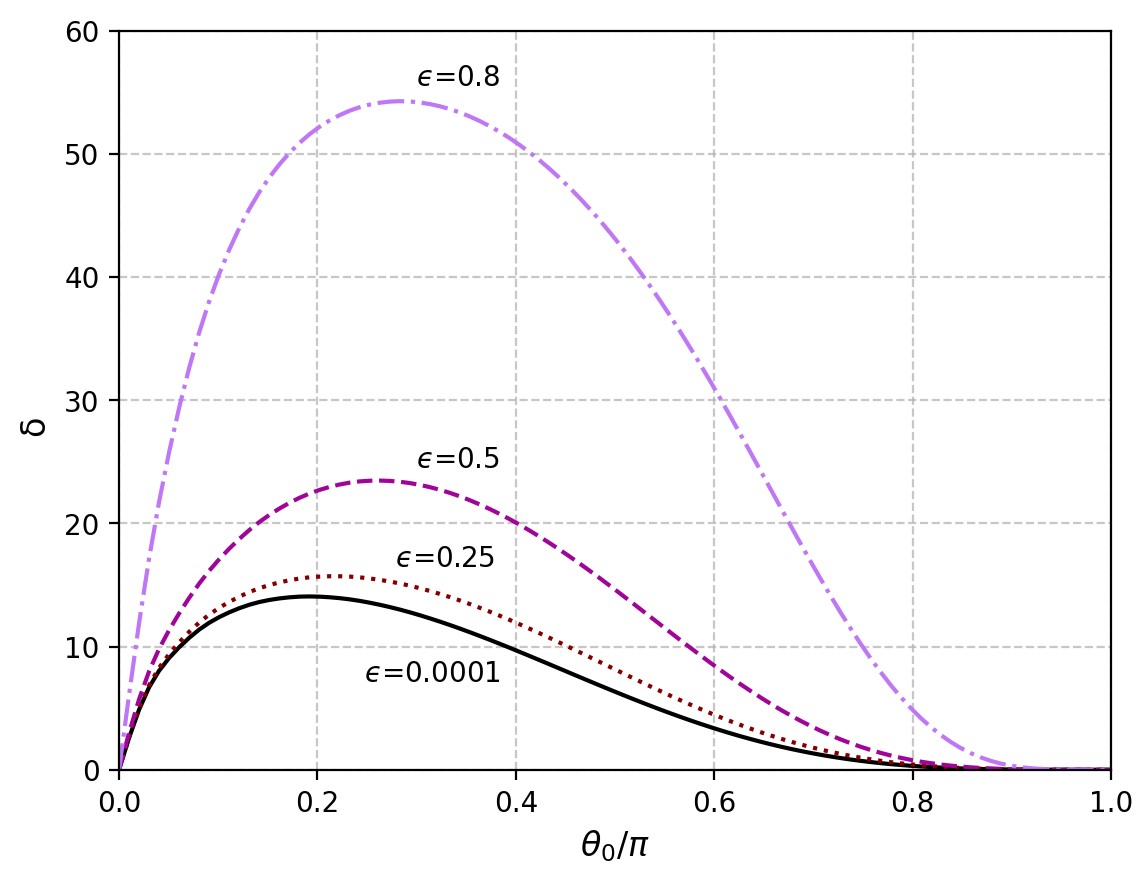}
	\caption{The relative percentage error $ \delta\left( \theta_0, \epsilon\right)$ (\ref{Err1}) the first approximation $J^{(1)}\left( \theta_0; \epsilon \right)$  (\ref{App2a}) as a function of the opening angle $\theta_0$ at four different magnitudes of the thickness ratio: $\epsilon = 10^{-4};\, 0.25;\, 0.5;\, 0.8$.}
	\label{fig5}
\end{figure*}
To present these discrepancies between $J$ and $J^{(1)}$ more clearly let us introduce the {\it relative percentage error} of the first approximation $J^{(1)}$ (\ref{App2a}) by the ratio 
\begin{align}
\delta \left( \theta_0; \epsilon \right) :=100 \left(J-J^{(1)} \right) /J \,. \label{Err1}  
\end{align}

Error (\ref{Err1}) is exhibited in Fig.~\ref{fig5}, where we plotted its dependence on the reactive size $\theta_0$ at different values of the thickness ratio $\epsilon $.

\begin{table}[h]
	\caption{Estimates of the maximum percentage error (\ref{Err1M}) of the first approximation $J^{(1)}$ (\ref{App2a}).}
	\label{tab:example_short}
	\begin{ruledtabular} 
		\begin{tabular}{lcc}
			$ \theta_0^{\ast} /\pi $ & $h $ & $ \delta_{M} \left(\theta_0^{\ast}, h \right) $ \\
			\colrule 
			0.191 & 0.9999  & 14.1 \\
			0.216 & 0.75 & 15.7 \\
			0.261 & 0.5 & 23.5 \\
			0.283 & 0.2 & 54.3 \\
		\end{tabular}
	\end{ruledtabular}
	\label{table3}
\end{table}
Due to continuity of the function (\ref{Err1}) let us define the maximum of the relative percentage error (\ref{Err1}) by the formula
\begin{align}
\delta_{M} \left(\theta_0^{\ast}, h \right):= \max_{\theta_0 \in (0, \pi)}\delta \left( \theta_0; 1-h \right)\,, \label{Err1M}  
\end{align}
where $\delta_{M} \left(\theta_0^{\ast}, h \right)$ is the monotonically decreasing function for $
h  \in  \left(0 , 1 \right) $. 

Specifically, Table~\ref{table3} presents the magnitudes $\delta_{M} \left(\theta_0^{\ast}, h \right)$ (\ref{Err1M}) associated some shell thicknesses parameter $h=1- \epsilon$ values in the range of small enough angular size of the active site $0.191 \le \theta_0^{\ast} /\pi \le 0.283$.

\subsection{Local concentration profile}~\label{subsec:profile}

In order to calculate the local concentration profile by means of formula (\ref{GenSol1b}) we have to keep so many terms of the series (\ref{GenSol1b}) (corresponding sequence elements $\{X_l\}$) to ensure required accuracy. 

\begin{figure*} [h!]
	\centering
	\includegraphics[width=0.94\textwidth]{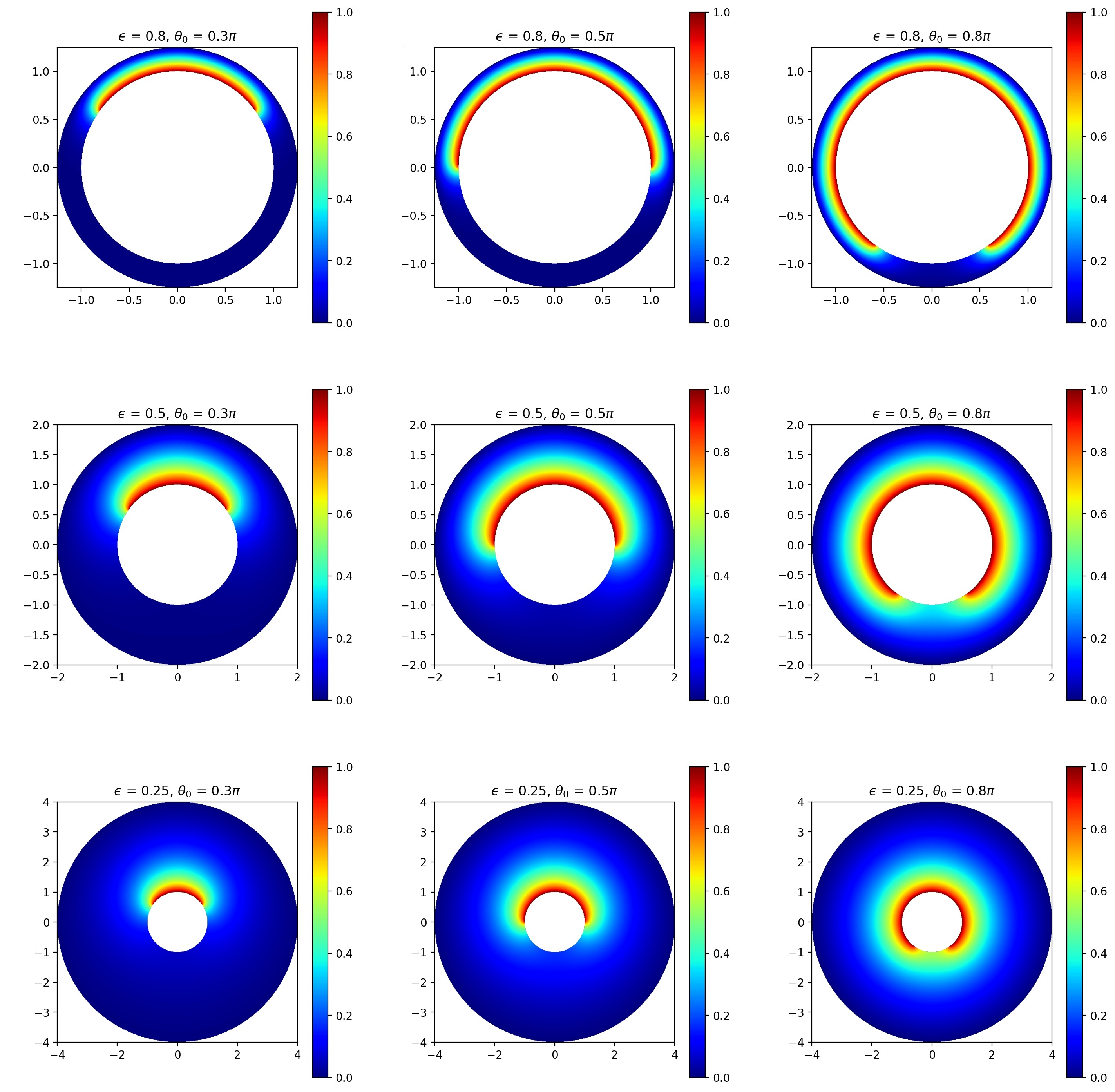}
	\caption{The maps of the local concentration  $u \left(\xi, \theta  \right) $ around the unit spherical RP (white)	governed by the expansion (\ref{GenSol1b}). Function $\hat u \left(\hat {x}, \hat {z} \right):= u \left(\xi\left(\hat {x}, \hat {z} \right), \theta \left(\hat {x}, \hat {z} \right) \right) $ was calculated within the main cross-section of the cavity in the RP dimensionless local Cartesian coordinates defined by Eqs. (\ref{Cross1}), (\ref{Cross2}). Pictures are given at fixed set of thickness ratios $\epsilon$ and angular sizes $\theta_0$. Color scale for concentration $u$ takes its values from the zero at the cell wall (dark blue) up to the unity at the active sites (dark red) with a fixed step value $0.2$. For better visibility of the plotted pictures we omit designations for both coordinate axes and concentration: $\hat {x}$, $\hat {z}$,  $\hat u \left(\hat {x}, \hat {z} \right) $.}
	\label{contours}
\end{figure*}

Fig.~\ref{contours} illustrates the obtained numerical results for the local concentration $u \left(\xi\left(\hat {x}, \hat {z} \right), \theta \left(\hat {x}, \hat {z} \right) \right) $ in the main cross-section of the system at issue using the RP local Cartesian coordinates  $\{ O; x, y, z\} $ (Fig.~\ref{fig1} (b)) rescaled by $R$. Thus, we carried out our calculations in the following dimensionless Cartesian coordinates: 
\begin{eqnarray}
\left\{ 0 < \hat {x}^2+\hat {z}^2 < 1/{\epsilon}^2 \right\} \cap \left\{\hat {y}=0  \right\}\,, \label{Cross1} \\ 
\left(\hat {x}, \hat {y}, \hat {z} \right):=\left(x/R, y/R, z/R \right)\,. \, \label{Cross2} 
\end{eqnarray}

Noteworthy is high rate of computations of desired concentration field $u \left(\xi, \theta \right) $. Really, increasing the truncation order by 2.5 times from 200 to 500 yields corrections of order $10^{-3}$, what does not affect the images resolution given in Fig.~\ref{contours}.

\section{Discussion and outline of approach}\label{sec:methodOut}

As might be expected on simple physical reasons both analytical approximation (\ref{App2a}) and results of numerical calculations given in Fig.~\ref{fig3} show that chemical anisotropy leads to a significant decrease of the diffusive interaction effects.

It is important to emphasize that in the present research we deal with rather intricate interplay of two competing subtle physical effects on the trapping rate $k$ or the rate correction factor $J$ (see Remark~\ref{synonyms}). 

Firstly Eq. (\ref{Seki6}) takes into account the effects due to anisotropic trapping (nonvanishing integral over the active site only):
\begin{itemize}
\item  Reduction of $k \left(\theta_0, \epsilon\right)$ due to anisotropic trapping by an active site $\partial \Omega_D \left(0<\theta <\theta_0 \right)$. 
\end{itemize}

Secondly, the trapping rate $k \left(\theta_0, \epsilon\right)$ depends on the  diffusive interaction between the RP and cavity wall because of conditions prescribed on the 2-connected boundary $\partial \Omega^+$ (defined by the union of $\partial \Omega_0$ and $\partial \Omega $ (\ref{Bound1a})):
\begin{itemize}
\item  Increasing of $k \left(\theta_0, \epsilon\right)$ due to the diffusive interaction between RP and the outer spherical cavity wall $\partial \Omega_0$ depending on the thickness ratio $\epsilon=R/R_0$.	 
\end{itemize}

It turned out that the theoretical method applied in the present paper may be entirely treated within the scope of the {\it generalized method of separation of variables} (GMSV) used previously to describe the diffusive interaction arising in many-sink reaction-diffusion systems.~\cite{Traytak24}

\begin{figure*} [h!]
	\centering
	\includegraphics[width=0.73\textwidth]{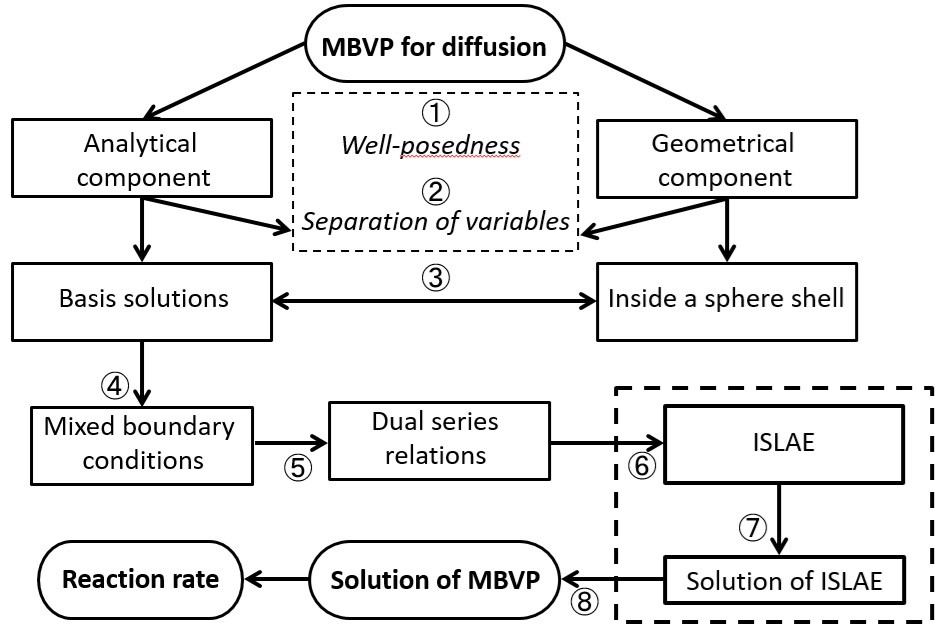}
	\caption{A block diagram of the generalized method of separation of variables for solution of the mixed boundary value problem (\ref{LE1})-(\ref{MBC2}) in the spherical shell domain $ \Omega^{+}_{\xi} $.}
	\label{block}
\end{figure*}

Summarising we conclude that the application of the aforementioned solution procedure to the Dirichlet-Neumann mixed boundary value problem (\ref{LE1})-(\ref{MBC2}) may be formulated as an algorithm, which in turn can be divided up into following eight main steps. 
\begin{enumerate}
\item Proof of the {\it well-posedness of the solution} to the posed mixed boundary value problem;~\cite{Volpert85}
\item  {\it Separation of variables}: decomposition of the desired solution in the form of the sum of partial solutions by means of the general linear superposition principle;
\item  Determination of the appropriate {\it basis solutions} (\ref{Psi1}), (\ref{Psi2}) to the mixed boundary value problem posed inside the {\it spherical shell domain $ \Omega^+ $} (\ref{Resc4}) with disconnected boundary;			
\item  Application of the basis solutions in order to satisfy the mixed boundary conditions (\ref{MBC1}), (\ref{MBC2});
\item Derivation of the {\it dual series relations} (\ref{DSR1a}), (\ref{DSR2a}); 
\item  Reduction of the dual series relations to a self-consistent {\it resolving infinite system of linear algebraic equations} (\ref{Bi28});
\item  Solution of the resolving infinite system of linear algebraic equations by means of functional analysis;	
\item Calculation of the local concentration profile and the microscopic reaction rate.
\end{enumerate}

Furthermore, for clarity sake, in Fig.~\ref{block} we have depicted the above full scheme of the GMSV application to the Dirichlet-Neumann mixed boundary value problem studied in the paper as a block diagram. 

Finally, let us make a few explanations concerning the suggested algorithm.

The second step means that the {\it linearity} of the original mixed boundary value problem is one of the primary requirements in the GMSV algorithm at issue.

Concerning the fifth step it should be noted the following. As is known from general theory of the {\it regular Sturm–Liouville problems}, application of the GMSV to the proper mixed boundary value problems for Laplace's equation, naturally, leads to discrete spectrum, that, in turn, leads to some kind of dual series with respect to eigenfunctions of the corresponding  Sturm–Liouville operator.~\cite{Ladyzhenskaya85}

The seventh step implies that the full investigation of the ISLAE solution including: existence, uniqueness, approximation, convergence etc. are developed in functional analysis.

Finally note that, quite often, an enzyme molecule has only one active site so, taking account this fact, above model seems not to be too oversimplified in order to elucidation the real situation.~\cite{Shoup81}

It should be stressed that the  Solc-Stockmayer active patches model is still remain basic for theoretical study of diffusion-influenced reactions occurring on the surface of immobile reactive centres in crowded or finite domains, particularly inside living cells (see, e.g., recent extensive research and survey given in Refs.~\onlinecite{Grebenkov25b,Bernoff25}).

\section{Concluding remarks}\label{sec:Conclusion}

In this paper we propose a generalization of the famous steady-state Berg's theory.~\cite{Berg93}   The ideally absorbed reactive particle was replaced by a particle with asymmetric patch-like reactivity. Then we managed to perform full mathematical study of the generalized Berg's theory and obtained both semi-analytical and numerical results.

It is common knowledge that problems posed in unbounded domains often appeared to be amenable to the analytical study including asymptotic approaches~\cite{Babenko09} but not for numerical ones. Numerical methods should be restricted by finite domains only. On the other hand mathematical models dealing with unbounded space are rather strong idealizations of real-world physical phenomena, since real physical processes occurring in finite, confined, crowded etc. domains.

Among obtained results the first approximation to the reaction rate (\ref{App2a}), being at the same time mathematically simple and physically lucid, seems to be very important. Indeed it can be served for understanding the coupling of anisotropic trapping and diffusive interaction effects taking place inside the cavity. This important question seems to be worthy of special attention.
Theoretical concepts of the effective steric factor and the rate correction factor were widely used by researchers studding of anisotropic chemical reactions and diffusive interaction effects, separately for many decades. However the study of the generalized Berg's model quite clearly showed that anisotropic chemical reactivity and diffusive interaction effects can be coupled. Moreover, as we have seen above these two important concepts were naturally unified within the scope of this work. 

In our opinion, however, another even more remarkable result was established in this study. 
We managed to find a close connection between the dual series relations method and the generalized method of separation of variables. The revealed  connection allowed us, similar to our previous work, divide the corresponding solution algorithm into eight basic and clear steps. This algorithm can be implemented  step by step in full detail to theoretical studding a number of future generalizations of the present simplified mixed boundary value diffusion problem. It should be particularly emphasized that using above formulated algorithm, we can find  solutions for a much wider class of important mixed boundary value diffusion problems. 

We have shown that the method of reduction may be used to solve numerically the resolving infinite system of linear equations to any degree of accuracy even when the simple-iteration method is not valid. Whereas the desired accuracy of the solution can be achieved just by increasing the truncation degree. Furthermore, note that the method has advantages such as shorter calculation time and more accurate results compared to known purely numerical and simulation methods. It is particularly surprisingly that the dual series
relations method requires the same computing time for both bounded and unbounded domains. Suggested powerful mathematical technique allows one to investigate rather sophisticated physical models to elucidate some fundamental issues in cell biology and related sciences, such as how the specific geometric configurations of reactants and boundaries with sterically specific reactivity the overall behavior of the reaction rates. Thus, following Ref.~\onlinecite{Eun17}, we can definitely conclude that for the one-patch Solc-Stockmayer model the dual series relations method gives the exact results.

Considered model is, of course, a gross simplification, nevertheless allows to elucidate the main features of the sterically specific diffusion-controlled reactions occurring on an active site of reactants placed into cell systems be studied. For instance, our results may be used to quantify rates of the trapping chemical reactions taking place in populations of isolated biological cells.

Future extension of the present work may include, e.g., generalizations which can take into account: bulk depletion of the $B$-particles, their rotational diffusion and time-dependent effects. 

And the last but not least point, in our forthcoming paper we intend to carry out the following research. Firstly, we will generalize the present theory to the case when the reactive particle and the cavity form a non-concentric spheres system. Secondly, we will compare our results against that obtained by means of the SLS approximation (constant flux method) in Ref.~\onlinecite{Eun17}.

\section*{Acknowledgements}
We sincerely thank Professor L. Dagdug and Professor G. Oshanin for drawing our attention to Ref.~\onlinecite{Dagdug24} and Ref.~\onlinecite{Oshanin24}.

\appendix

\section{Some physical and mathematical assumptions and notions} \label{sec:appendix}

In this Appendix, for the readers’ convenience, we present some basic physical assumptions of Smoluchowki's  theory along with some mathematical notation, definitions which are used throughout the paper.

\subsection{Physical background}  \label{subsec:Backappendix}

Here we just briefly summarize the physically motivated assumptions underlying the Smoluchowski theory.

(1) Standard treatments of the theory of diffusion-controlled reactions assume that the host medium is unbounded, quiescent, isotropic and homogeneous, therefore
mathematically it is well modeled by the common 3D  {\it Euclidean space $ {\mathbb R}^3 $}. Thus, Euclidean space $ {\mathbb R}^3 $ may be called the {\it physical space}.

(2) Suppose that the following properties of reactants  $A$ and $B$ included in the reaction scheme (\ref{Zv00}) hold true.
\begin{itemize}
\item
Both reactants are neutral (uncharged) particles.	
\item  
Reactants $B$ are identical non-interacting between each other and very small (point-like) particles.	 
\item  
Reactants $A$ are spherical particles of infinite capacity.
\item 
Reactants $A$ are much larger than the $B$-particles, and therefore they can be regarded as immobile.
\item 
Generally speaking particles $A$ are chemically asymmetric, whereas  $B$-particles have isotropic reactivity.
\end{itemize}

(3) Diffusion of $B$-particles can be described by a real-valued function that satisfies an appropriate diffusion equation.

(4) The specific form of the boundary conditions prescribed on the reactants surfaces are completely determined by the features of reaction-diffusion processes under studying. 

\begin{definition} \label{confinement}
Under {\it geometrical confinement}, contrary to the {\it confinement} arising due to, e.g., action of some dynamic potentials, we understand belonging to a bounded (finite) domain of a physical space.
\end{definition} 

\begin{definition} \label{ActiveS}
Any region of the reactive particle that binds to the chemically isotropic $B$-particles is called the fully absorbed {\it active site}. Provided it belongs to the surface of the RP it is also called active site of patch type or just {\it active patch}.
\end{definition} 

We would like to draw the reader's attention to the simultaneous use in literature terms "active" and "reactive" with respect to the site (patch).

\subsection{Basic mathematical notations and definitions}\label{subsec:Basic}

Throughout this paper we will use the common mathematical symbols: $ \Bbb{N} $ denotes the set of natural numbers; $ \Bbb{R} $ the reals, $ \Bbb{R}_+:= \lbrace x \in \Bbb{R}: 0< x < + \infty\rbrace $  the strictly positive reals;  and $ \Bbb{R}^n $ the nD vector space associated with nD Euclidean space comprising points $\mathbf{r}:= \left(x_1,... , x_n \right)$  with respect to an origin $O$ such that $\Bbb{R}^1=\Bbb{R}$.

Symbol $\overline{l, m}$ means that all integers (including $\lbrace \infty \rbrace $) from $ l $ to $ m $ are taken their values successively.

As is customary, let $\partial\Omega$ denotes the closed boundary of a domain $\Omega \subset  \Bbb{R}^n $ (connected
open subset), such that its closure is $\overline{\Omega}:={\Omega}\cup \partial\Omega$. 

Notation $ g\left(a \pm \right):=\lim_{ \epsilon (>0) \to 0 }g\left(a \pm \epsilon\right)$ stands for the one-sided limits of a real-valued function $ g\left(x \right) $ at a point $ a \in \Bbb{R} $.

The {\it restriction of a function $f : \Omega \to \R$ on a boundary subset} $ \partial\omega \subset \partial \Omega$
is usually denoted by $\left. f\right| _{\partial\omega}$ and called the {\it boundary function at the boundary part $\partial\omega$ }.

It is common knowledge that one-to-one mapping of Cartesian coordinates to spherical coordinates: $ \{ O; x, y, z\} \mapsto \left\{O; r, \theta ,\phi \right\} $ in domain $ \Omega^- $ defined by (\ref{Dom1}) exists if we exclude $Oz$ axis and half-plane $ \lbrace x \geq 0, y =0 \rbrace $.~\cite{Traytak24} Thus we have known relations
\begin{align}
x= r \sin \theta \cos \phi\,, \quad y= r \sin \theta \sin \phi\,,\quad z=r\cos \theta 	\label{SphC1}
\end{align} 
under conditions:  $ \lbrace 0< r=\sqrt{x^2+y^2+z^2} <+ \infty\rbrace $,   $\lbrace 0 < \theta < \pi\rbrace $ and  $\lbrace 0 < \phi < 2\pi\rbrace $.

Mostly a mathematical model of physical processes may be formulated with respect to a real-valued function given in a domain $\Omega \subset \Bbb{R}^n$ (i.e.  $u:\Omega \to \Bbb{R} $) and its restriction on the  associated boundary values. Function $u$ is a solution to some partial differential equation. So to formulate the reaction-diffusion problem rigorously, we need to consider the {\it analytical components} of the problem (differential equations and relations), while domain and its boundaries are called {\it geometrical components}.~\cite{Rvachev95}

\subsection{The mixed boundary conditions}\label{subsec:MixedBVP}

Since, despite more than a century of its history the terminology on the mixed boundary value problems is still  not well established, we dwell on this subject.

First, let us reformulate the known definition of a domain partition~\cite{Repin08} in the form used in the paper.
\begin{definition} \label{Partition1}
For the domain $\Omega$ boundary $\partial\Omega$ a family of its $m$ open subsets $  \lbrace \partial \Omega_1, \partial\Omega_2, ..., \partial\Omega_m \rbrace $ ($\partial\Omega_i \subseteq \partial\Omega $ when $i=\overline{1, m}$) is termed the {\it $m$-partition of boundary $ \partial\Omega $}, provided for this family the following conditions hold true:
\begin{eqnarray}
	\partial\Omega =  \bigcup\limits_{i=1}^m \partial \Omega_i   \,, \label{Part1}	\\	
	\partial \Omega_i \cap \partial \Omega_j = \varnothing \quad \mbox{if} \quad i\neq j\, .  \label{Part2}
\end{eqnarray}
\end{definition}
Any element $ \partial \Omega_i $ in the union (\ref{Part1}) is called a {\it connected component of boundary} $ \partial \Omega $. 

Provided $ \Omega $ is a 3D bounded domain with a smooth boundary $ \partial\Omega $ there exists its $m$-partition (\ref{Part1}). Note that mostly in biophysical applications a bounded domain boundary is partitioned into two part: absorbing and reflecting.~\cite{Schuss13} 

Two types of mixed boundary conditions, which may be imposed for a given partition are distinguished.~\cite{Price07}

\begin{definition}\label{defin2}
If the boundary conditions differ on different parts of a given connected component of a boundary $ \partial \Omega_i$ these conditions are referred to as {\it proper mixed boundary conditions}.
\end{definition}
Physically the proper mixed boundary condition means that a given connected component $ \partial \Omega _i $ is characterized by a prescribed heterogeneous reactivity. 
\begin{definition}\label{defin3}
Provided different boundary conditions are posed on different connected components (\ref{Part2}) they are termed {\it improper mixed boundary conditions}.
\end{definition}
Hence the appropriate boundary value problems under proper and improper mixed boundary conditions are called {\it proper} and {\it improper mixed boundary value problems}, respectively. As we have pointed out above in this paper we studied only proper mixed boundary value problems. 
\begin{remark} \label{remarkCD}
It should be highly stressed here that the difference in types of boundary conditions plays a crucial role in the above definition.  
Say, the Dirichlet conditions with different values of boundary functions $f_i$ imposed on the corresponding connected components  $ \partial \Omega_i$ of the boundary  (that is $\left. u\right| _{\partial \Omega_i}=f_i$) are not mixed boundary conditions all by any means (see, e.g., problem 176 in p. 90 in Ref.~\onlinecite{Lebedev79}).
\end{remark}

\subsection{Condition on the edge}~\label{subsec:profile}

We currently know that for the uniqueness of the solution to the mixed problems, it is necessary to specify the so-called {\it condition on the edge} $\gamma$ (see Sec.~\ref{sec:Statement}), where the smoothness of the solution may be violated. The cause of this kind of behavior could be happened due to violation of the compatibility condition between Dirichlet and Neumann boundary conditions at the edge.

It is important to emphasize the singular behavior of the field $\rho \left( {\bf r}\right)$ and its normal derivative at a vicinity of the edge $\gamma$. Since the singularity at the edge appeared to be integrable it does not effect the calculations of the reaction rate. However, one should keep in mind that difficulties may arise during calculation of the local concentration profile $\rho \left( {\bf r}\right)$ in a vicinity of the edge (see Subsection~\ref{subsec:profile}).
Edge condition should be imposed to make solution unique~\cite{Vinogradov02} 
\begin{align}
\int_{\Omega_{\gamma}} \Vert \bm{\nabla}\rho \Vert^2 d{\bf r} < +\infty \label{Edge1}  
\end{align} 
such that $\gamma \subset \Omega_{\gamma} \subset \overline{\Omega}^+ $.

Note in passing that similarly to a technique proposed in Ref.~\onlinecite{Babaylov85} one can show that integral condition (\ref{Edge1}) may be replaced by
\begin{align}
\lim_{\theta \to \theta_0-}\left. \frac{\partial \rho}{\partial r}\right| _{r =R+}=0 \label{Edge2}  
\end{align} 
however, this matter is beyond the main subject of our paper.


\bibliography{Liter}


\end{document}